\documentclass[%
 aip,
 amsmath,amssymb,
 reprint,%
]{revtex4-1}

\usepackage{graphicx}
\usepackage{dcolumn}
\usepackage{bm}

\usepackage[utf8]{inputenc}
\usepackage[T1]{fontenc}
\usepackage{mathptmx}
\usepackage{etoolbox}

\usepackage[version=4]{mhchem}
\usepackage{chemformula}

\makeatletter
\def\@email#1#2{%
 \endgroup
 \patchcmd{\titleblock@produce}
  {\frontmatter@RRAPformat}
  {\frontmatter@RRAPformat{\produce@RRAP{*#1\href{mailto:#2}{#2}}}\frontmatter@RRAPformat}
  {}{}
}%
\makeatother
\begin{document}

\preprint{AIP/123-QED}

\title{Tumor-induced lipid loss physically accelerates breast cancer invasion into mammary adipose tissue}

\author{Garrett F. Beeghly}
\thanks{These authors contributed equally to this work.}
\affiliation{Nancy E. and Peter C. Meinig School of Biomedical Engineering, Cornell University, Ithaca, New York 14853, USA\looseness=-1}

\author{Dong Wang}
\thanks{These authors contributed equally to this work.}
\affiliation{Department of Mechanical Engineering, Yale University, New Haven, Connecticut 06520, USA\looseness=-1}

\author{Bo Ri Seo}
\affiliation{Nancy E. and Peter C. Meinig School of Biomedical Engineering, Cornell University, Ithaca, New York 14853, USA\looseness=-1}

\author{Yitong Zheng}
\affiliation{Department of Mechanical Engineering, Yale University, New Haven, Connecticut 06520, USA\looseness=-1}

\author{Joseph E. Druso}
\affiliation{Nancy E. and Peter C. Meinig School of Biomedical Engineering, Cornell University, Ithaca, New York 14853, USA\looseness=-1}

\author{Benjamin D. Hopkins}
\affiliation{Englander Institute for Precision Medicine, Weill Cornell Medicine, New York, NY 10021, USA\looseness=-1}

\author{Linda T. Vahdat}
\affiliation{Dartmouth Health and Department of Medicine, Dartmouth Geisel School of Medicine, Hanover, New Hampshire 03755 USA\looseness=-1}

\author{Neil M. Iyengar}
\affiliation{Winship Cancer Institute at Emory University, Atlanta, GA 30308, USA\looseness=-1}

\author{Mark D. Shattuck}
\affiliation{Benjamin Levich Institute and Physics Department, City College of New York, New York, New York 10031, USA\looseness=-1}

\author{Corey S. O'Hern}
\email{corey.ohern@yale.edu}
\affiliation{Department of Mechanical Engineering, Yale University, New Haven, Connecticut 06520, USA\looseness=-1}
\affiliation{Integrated Graduate Program in Physical and Engineering Biology, Yale University, New Haven, Connecticut 06520, USA\looseness=-1}
\affiliation{Department of Physics, Yale University, New Haven, Connecticut 06520, USA\looseness=-1}

\author{Claudia Fischbach}
\email{cf99@cornell.edu}
\affiliation{Nancy E. and Peter C. Meinig School of Biomedical Engineering, Cornell University, Ithaca, New York 14853, USA\looseness=-1}
\affiliation{Kavli Institute at Cornell for Nanoscale Science, Cornell University, Ithaca, NY 14853, USA\looseness=-1}


\date{\today}

\begin{abstract}

Obesity is a major risk factor for breast cancer, yet how obesity-associated changes in the physical properties of white adipose tissue (WAT) influence tumor invasion remains poorly understood. Here, we combine experimental and computational approaches to investigate this question. Using mouse models and human mastectomy samples, we show that obesity not only increases adipocyte size, but also asphericity. We observe that tumor-conditioned media drives lipid loss and dedifferentiation of adipocytes into myofibroblast-like cells\textit{ in vitro }and that adipocytes in high-fat diet-fed mice lose lipid at a faster rate than those in normal diet-fed mice in response to mammary tumors \textit{in vivo}.  We then develop discrete element method (DEM) simulations to understand how these obesity-induced changes in cell and tissue properties alter breast cancer invasion under high-fat and normal diet conditions. In DEM simulations, adipocytes are modeled as deformable polygons and polyhedra, and cancer cells are modeled as soft, adhesive disks and spheres in two and three dimensions, respectively. Invasion is driven by cancer cell proliferation and tumor-induced lipid loss is modeled as corresponding decreases in adipocyte size. DEM simulations indicate that obesity-associated increases in tissue pressure induce adipocyte deformation, consistent with elevated asphericity in experimental data. At small lipid loss rates, the degree of cancer invasion is only weakly affected by tissue pressure. However, at sufficiently large lipid loss rates, we find that tissue pressure accelerates cancer invasion. Together, these results suggest that lipid loss physically remodels WAT to facilitate breast cancer invasion, and that obesity may exacerbate this effect.

\end{abstract}

\maketitle

\section*{Introduction}

Breast cancer originates in the mammary glands and has favorable patient survival for early-stage disease, but poor prognosis when tumors metastasize~\cite{harbeck2019natreview}. Notably, the frequency of metastasis and mortality rate of breast cancer increase with obesity~\cite{calle2003nejm, renehan2008lancet, protani2010}. Therefore, it is crucial to understand and prevent the initial invasion of breast cancer cells into the white adipose tissue (WAT) surrounding the mammary glands, especially under obese conditions. Most prior work has focused on how endocrine, inflammatory, and metabolic dysfunction of obese WAT impacts tumor progression~\cite{park2011endorev, khandekar2011natrevcancer, hopkins2016jco}, but altered physical properties of the local tissue microenvironment are also important \cite{beeghly2022review, beeghly2023review}. Indeed, previous studies have shown that the mechanics of cancer cells and surrounding extracellular matrix regulate where and how cancer invasion occurs \cite{ilina2020natcellbio, cross2007natnanotechnol, rianna2020mbc, provenzano2006bmcmed, levental2009cell}. 

Yet compared to tumor and ECM mechanics, how the physical properties of mammary WAT impact breast cancer invasion remains poorly understood. WAT exhibits a distinct tissue morphology characterized by packings of large, spherical adipocytes. Moreover, adipocytes are not directly adherent to each other. Instead, collagen fibers in the interstitial regions between adipocytes provide WAT with its mechanical stability \cite{sakers2022cell, ghaben2019natrevmcb}. Tumor cells must physically navigate this topography during the initial stages of metastasis~\cite{knode2026cellbiomater} and the physical properties of WAT can also change under various pathological conditions. For example, obesity induces volumetric expansion of the adipocytes~\cite{ghaben2019natrevmcb} and fibrotic remodeling of interstitial collagen~\cite{seo2015scitransmed}, while tumor-derived factors cause adipocytes to lose lipid and decrease in size~\cite{dirat2011cancerresearch}. Understanding how the dynamic physical properties of WAT (i.e. changes to adipocyte size and shape) modulate tumor invasion will help establish new physical biomarkers that can identify patients most susceptible to invasive disease.

While tuning the mechanics of WAT remains impractical with conventional cell culture and animal models, computational approaches can provide control over individual tissue parameters. However, many current approaches to simulate cellular systems are not able to capture important physical features of WAT, such as cell deformability and the absence of direct cell-cell adhesion. For example, coupled reaction-diffusion equations treat tumor cells and tumor-derived factors as diffusive chemical species~\cite{anderson2000theormed, anderson2006cell, araujo2004bullmathbio}, but do not have the ability to quantitatively describe cell shape and mechanics. Tessellation-based models~\cite{bi2015naturephys, bi2016prx, sussman2018softmatter} are constrained to confluent tissues with direct cell-cell adhesions, while tumor invasion is largely driven by tumor-stroma interactions~\cite{anderson2000theormed} and the loss of cell-cell adhesions during the epithelial-mesenchymal transition~\cite{ilina2020natcellbio}. Lattice Boltzmann approaches~\cite{graner1992prl} can model non-confluent tissues and single-cell properties, but these simulations are computationally costly~\cite{bresler2019epje} and often assume that the cell interior is fluid-like, which does not properly capture the viscoelastic and plastic mechanical behavior of cells ~\cite{kasza2007curropiocellbio}.

Instead, we leverage the recently developed deformable particle model (DPM)~\cite{boromand2018prl, treado2021prmater, wang2021softmatter, treado2022jrsi, zheng2024aplbio}, which can quantitatively recapitulate cell shape and mechanics in confluent and non-confluent tissues. We perform discrete element method (DEM) simulations of cancer cells invading WAT in both two dimensions (2D) and three dimensions (3D). By including 2D systems, we can more readily study the effects of confinement on cancer invasion and compare our simulation results to histological cross-sections. Notably, we model WAT as dense packings of deformable polygons (polyhedra) in 2D (3D), whose centers of mass are tethered to a virtual substrate to mimic adhesion to the ECM. Cancer cells are modeled as soft, cohesive circular (spherical) particles in 2D (3D) since the magnitude of cancer cell shape fluctuations is much smaller than that of adipocytes. Cancer cells and WAT are de-mixed when the simulations are initialized, with a smooth interface between them. Cancer cells then proliferate at this interface, which drives tumor invasion into the surrounding adipose tissue. During the simulations, we quantify the degree of invasion by calculating the interfacial length (area) in 2D (3D) shared by cancer cells and adipocytes as a function of time. Collectively, these simulations allow us to study tumor invasion as a function of cancer cell-cell adhesion, tissue pressure, as well as adipocyte size, shape, and tethering to the ECM.

Here, we first compare the size and shape of adipocytes under obese conditions in mice and human samples and investigate how breast cancer cells invade WAT in mice pre-fed a normal versus high-fat diet. Using DEM simulations, we then show how elevated tissue pressure, induced by lipid accumulation, deforms adipocytes in obese conditions, but has little impact on baseline cancer invasion. We also examine how the pinning strength of adipocytes to the ECM influences the extent and phenotype of invasion. Experimentally, we further demonstrate that cancer cells induce adipocyte lipid loss by comparing adipocytes in the tumor versus stroma, and by exposing \textit{in vitro} differentiated adipocytes to tumor-derived factors. In DEM simulations, we show how increased lipid loss, coupled with greater tissue pressure, can synergistically promote cancer invasion in obese conditions—potentially contributing to the higher morbidity and mortality rates associated with obesity. Taken together, our results suggest that, in addition to promoting breast cancer cell proliferation and migration directly \cite {beeghly2026cellrep}, tumor-mediated lipid loss induces physical changes in WAT that increase interstitial void space and reduce the mechanical barrier to invasion. Moving forward, interfering with the ability of tumor cells to remodel their physical microenvironment could be exploited as a new approach to limit cancer dissemination.

\begin{figure*}[t!]
    \centering
    \includegraphics[width=\linewidth]{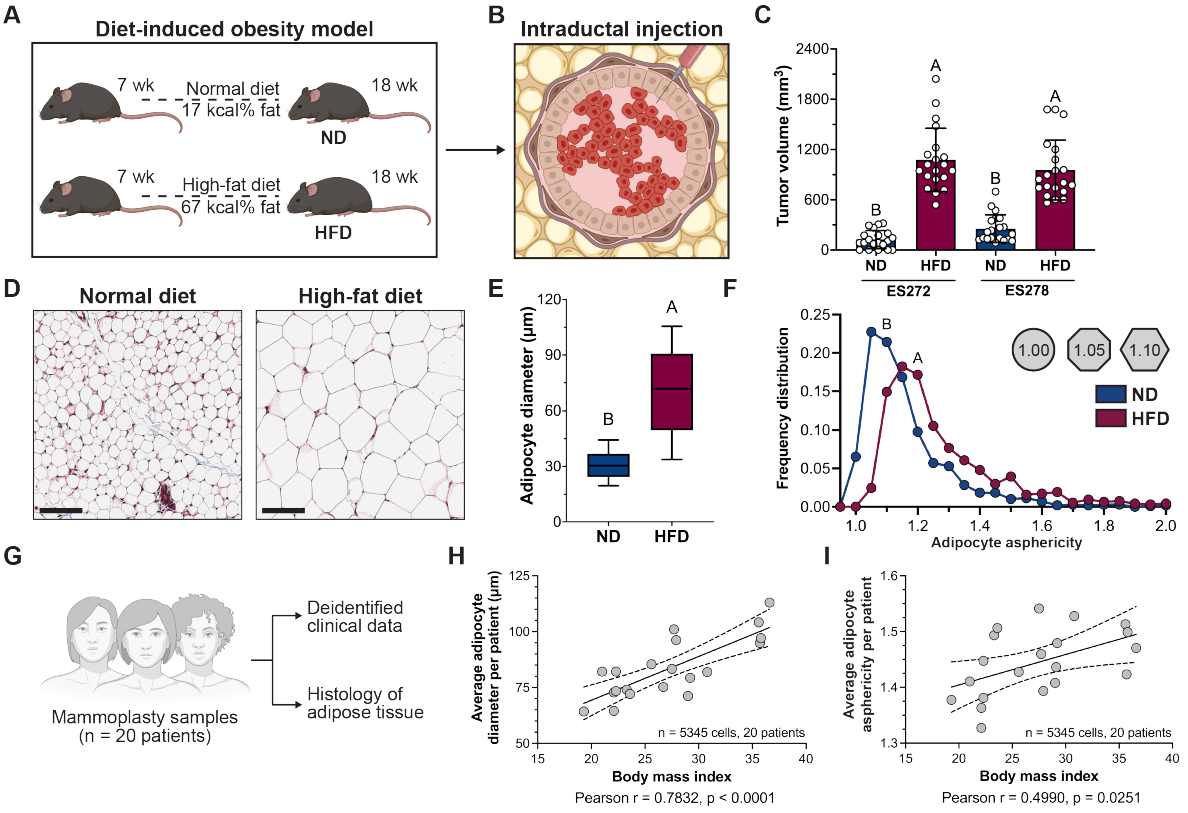}
    \caption{\textbf{Obesity increases adipocyte size and asphericity in mice and humans.} (A) Schematic of diet-induced obesity in C57BL/6 mice. (B) Schematic of the orthotopic breast cancer model via intraductal injection of syngeneic ES272 or ES278 cells. (C) Endpoint tumor volumes in mice pre-fed a normal diet (ND) or a high-fat diet (HFD) for $11$ weeks prior to intraductal injection. (D) Representative sections of mammary adipose tissue in ND and HFD mice. Scale bars = $100~\mu m$. (E) Box plot distributions of adipocyte diameter in ND and HFD mice. (F) Frequency distributions of adipocyte asphericity in ND and HFD mice. Example shapes with corresponding values of asphericity are provided for reference. (G) Schematic of human adipose tissue collection from patients undergoing mastectomy. (H) Correlation between body mass index (BMI) and average adipocyte diameter per patient. (I) Correlation between body mass index (BMI) and average adipocyte asphericity per patient. All data are presented as mean $\pm$ one standard deviation unless otherwise noted. Statistics were performed with a one-way analysis of variance (ANOVA) with Tukey's multiple comparisons test (C), Mann-Whitney U test (E, F), or Pearson correlation test (H, I). Compact letter display indicates statistical significance. Groups not sharing a letter are significantly different ($p < 0.05$).}
    \label{fig:obesity_exp}
\end{figure*}

\section*{Results}

\subsection*{Obesity increases adipocyte size and asphericity in mice and humans} 

To assess how the physical properties of mammary WAT change under obese conditions and impact breast cancer invasion, we used a diet-induced model of murine obesity~\cite{kleinert2018natrevendo, seo2015scitransmed, naftaly2022ijms}. In this model, female C57BL/6 mice were fed a normal diet (ND, $17$ kcal\% fat) or a high-fat diet (HFD, $67$ kcal\% fat) for $11$ weeks (Fig.~\ref{fig:obesity_exp}A) prior to intraductal injection of ES272 or ES278 syngeneic breast cancer cells (Fig.~\ref{fig:obesity_exp}B). We first validated that both ES272 (PI3KCA H1047R) and ES278 (PI3KCA H1047R, MYC overexpression) tumors grew substantially larger in mice pre-fed a HFD compared to mice pre-fed a ND as expected (Fig.~\ref{fig:obesity_exp}C). To then assess how the different diet conditions affect mammary adipocytes, we subjected tumor-adjacent WAT to histological staining (Fig.~\ref{fig:obesity_exp}D). Visibly, WAT from HFD mice had larger and more polygonal adipocytes compared to ND mice. We quantified the distribution of cell sizes across conditions, finding the average diameters to be $31 \mu$m and $72 \mu$m in ND and HFD mice, respectively (Fig.~\ref{fig:obesity_exp}E). Likewise, we also measured differences in the asphericity of the adipocytes, ${\cal A}^{2D} = p^2/(4 \pi a)$, where $p$ and $a$ are the perimeter and area of each adipocyte cross-section. For example, a circle has an asphericity of ${\cal A}^{2D} = 1$ and ${\cal A}^{2D}$ increases as adipocyte cross-sections deviate from a circle. We found a significant rightward shift in the asphericity distribution of adipocytes in HFD mice versus ND mice, with average values of ${\cal A}^{2D} \approx 1.29$ and $1.17$, respectively (Fig.~\ref{fig:obesity_exp}F). Given that adipocytes are spherical in isolation (and thus circular in 2D cross-sections)~\cite{beeghly2026cellrep}, these findings suggest that HFD WAT exhibits increased mechanical stress, which induces adipocytes to deform and deviate from ${\cal A}^{2D}=1$. To determine if similar changes occur in response to obesity in humans, we analyzed mammary adipocytes from a cohort of $20$ women undergoing mastectomy (Fig.~\ref{fig:obesity_exp}G). Similar to the results in mice, we observed a strong positive correlation between the average diameter of adipocytes and body mass index (BMI) (Fig.~\ref{fig:obesity_exp}H). We also found a positive correlation between the average asphericity of adipocytes and BMI (Fig.~\ref{fig:obesity_exp}I). Our data indicate that obesity alters the physical features of mammary WAT by increasing adipocyte size and asphericity in both mice and humans.

\subsection*{Adipose tissue pressure modulates adipocyte cell shape but not breast cancer invasion}

\begin{figure*}[t!]
    \centering
    \includegraphics[width=\linewidth]{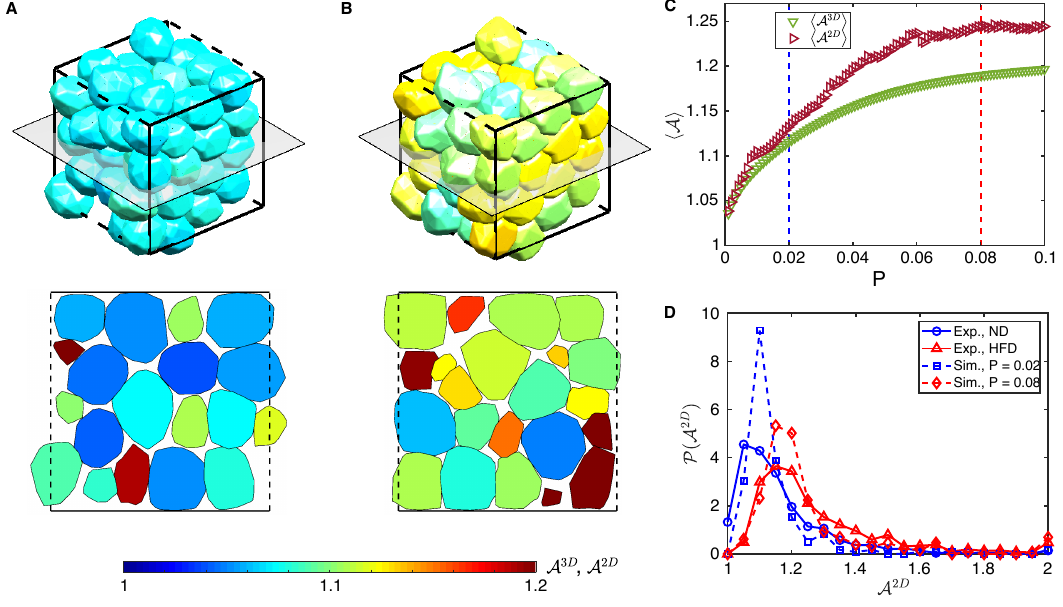}
    \caption{\textbf{Discrete element method simulations of adipose tissue provide insight into cell shape changes as a function of tissue pressure.} (A) Top: Adipose tissue at pressure $P = 0.02$, where adipocytes are modeled as deformable polyhedra. Bottom: Adipocyte cross-sections for the tissue in the top image at the location indicated by the gray square. (B) Top: the same adipose tissue as in panel (A), but at pressure $P = 0.08$. Bottom: adipocyte cross-sections for the tissue in the top image at the location indicated by the gray square. In both panels (A) and (B), the color indicates adipocyte asphericity. (C) Average 3D asphericity for adipocytes in adipose tissue ($\mathcal{A}^{3D}$, green downward triangles) and average asphericity for random cross-sections of the same adipose tissue in 2D ($\mathcal{A}^{2D}$, purple rightward triangles), plotted versus tissue pressure $P$. Blue and red dashed vertical lines correspond to the tissue pressures at which $\mathcal{A}^{2D}$ matches the values observed in ND and HFD mice as shown in Fig.~\ref{fig:obesity_exp}F, respectively. (D) Probability distribution of the 2D asphericity $\mathcal{P}(\mathcal{A}^{2D})$ of adipocyte cross-sections at pressures indicated by the dashed lines in panel (A), as well as those in ND and HFD mice in Fig.~\ref{fig:obesity_exp}F. }
    \label{fig:cellshape}
\end{figure*}

Given that adipocytes are larger (Fig.~\ref{fig:obesity_exp}E) and more non-spherical (Fig.~\ref{fig:obesity_exp}F) in HFD mice compared to ND mice, we hypothesize that adipose tissue pressure is greater in HFD mice due to increased lipid content. To test this hypothesis, we numerically generated static packings of adipocytes as a function of tissue pressure $P$ in 3D DEM simulations, as shown in Figs.~\ref{fig:cellshape} A-B. We initially focused on packings of adipocytes with the same diameter to decouple the effects of cell size from tissue pressure. To compare the results of the DEM simulations to experimental observations, we generated cross-sections of the simulated adipocytes (Figs.~\ref{fig:cellshape} A-B). Notably, we observed that adipocytes become more polygonal and that the spacing between adipocytes decreases with increasing pressure (Figs.~\ref{fig:cellshape} A-B). We find that the asphericity of adipocytes both in 3D $\mathcal{A}^{3D}$ and in 2D cross-sections $\mathcal{A}^{2D}$ first increases with $P$ and then reaches a plateau when $P$ is greater than $\sim 0.07$ (in units of the bulk modulus of the adipocytes) as shown in Fig.~\ref{fig:cellshape}C. The plateau corresponds to the regime where the packing of adipocytes becomes confluent and their shape parameters can no longer increase. This relationship between $P$, $\mathcal{A}^{3D}$, and $\mathcal{A}^{2D}$ supports our initial hypothesis that increased tissue pressure, driven by lipid accumulation in individual adipocytes, contributes to greater adipocyte deformation in HFD mice. Furthermore, we find that the probability distributions of $\mathcal{A}^{2D}$ at $P = 0.02$ and $P = 0.08$ from the DEM simulations match well with experimental values obtained in ND and HFD mice, respectively (Fig.~\ref{fig:cellshape}D), supporting the use of DEM simulations to study the physical properties of adipocytes under physiological conditions. 

\begin{figure*}[t!]
    \centering
    \includegraphics[width=\linewidth]{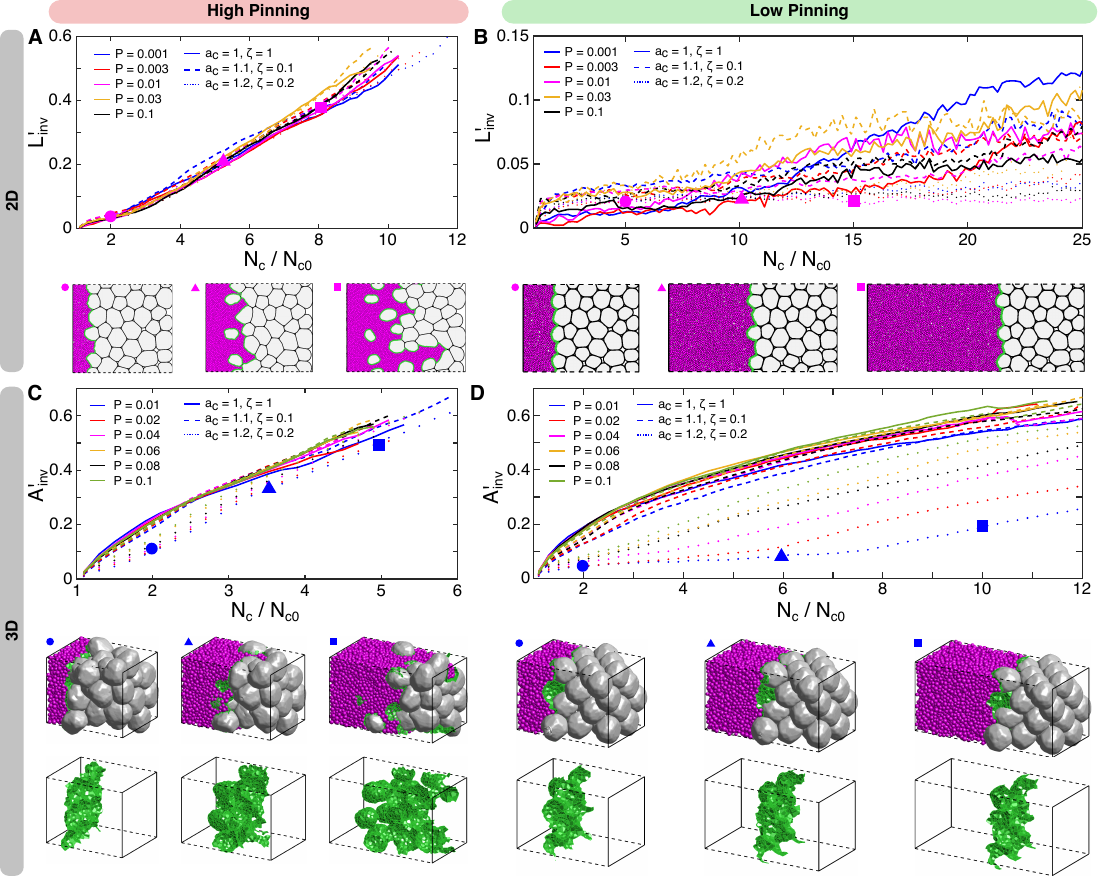}
    \caption{\textbf{Discrete element method simulations indicate that cancer invasion varies with adipocyte pinning strength but not tissue pressure.} (A-B) Breast cancer invasion in 2D with different adipocyte pinning strengths: (A) $K_{pin} = 0.2$ and (B) $K_{pin} = 0$. Top: degree of cancer invasion $L_{inv}'$ plotted versus the number of cancer cells rescaled by the initial value, $N_c/N_{c0}$ at different pressures $P$, degrees of cancer cell-cell adhesion $a_c$, and depth $\zeta$ from Eq.~\ref{eq:energy_cc}. Bottom: Snapshots from the DEM simulations for $P = 0.01$, $a_c = 1.2$, and $\zeta = 0.2$ shown in the top panel at the values of $N_c/N_{c0}$ indicated by the symbols. (C-D) Breast cancer invasion in 3D with different pinning strengths: (C) $K_{pin} = 0.2$, and (D) $K_{pin} = 0$. Top: Degree of cancer invasion $A_{inv}'$ plotted versus the rescaled number of cancer cells $N_c/N_{c0}$ at different $P$, $a_c$, and $\zeta$. Middle: Snapshots from the DEM simulations at pressure $P = 0.01$, cancer cell adhesion $a_c = 1.2$, and $\zeta = 0.2$ as shown in the top panel at the values of $N_c/N_{c0}$ indicated by the symbols. Bottom: The interface between cancer cells and adipocytes for the snapshots above. In (A)-(D), cancer cells are magenta, adipocytes are gray, and the interface between cancer cells and adipocytes is green.}
    \label{fig:cellinv_sim}
\end{figure*}

We then performed DEM simulations of proliferation-driven breast cancer invasion into WAT in both 2D and 3D over a range of tissue pressure $P$, cancer cell-cell adhesion (both range and strength), and adipocyte pinning to the ECM, without changing adipocyte size during invasion. At high pinning strength, cancer invasion into WAT proceeds as cancer cells mix with the adipocytes (Figs.~\ref{fig:cellinv_sim} A and C) and the degree of invasion does not depend strongly on $P$ in 2D (Figs.~\ref{fig:cellinv_sim}A) and 3D (Fig.~\ref{fig:cellinv_sim}C). In particular, the same number of cancer cells ($N_c/N_{c0}$) is required to reach a given degree of invasion ($L_{inv}'$ in 2D and $A_{inv}'$ in 3D) regardless of $P$. The independence of invasion on $P$ suggests that tissue pressure alone does not cause the enhanced tumor growth observed in the HFD mouse model (Fig.~\ref{fig:obesity_exp}C), highlighting the need to investigate other biophysical factors that promote cancer invasion. 

Cancer cell-cell adhesion has also been shown to impact cell migration speed and phenotype~\cite{ilina2020natcellbio}.  However, we find that invasion is not strongly affected by increased cancer cell-cell adhesion in 2D simulations (Fig.~\ref{fig:cellinv_sim}A). In contrast, in 3D simulations, increased cancer cell-cell adhesion impedes invasion. We attribute the limited effect of cell-cell adhesion on invasion in 2D to spatial confinement. In 2D, cancer cells with or without adhesion can be easily blocked by obstacles since creating a percolating network of void space is more difficult.  In addition, we find that the phenotype of cancer invasion is mainly determined by the strength of adipocyte pinning to the ECM. With large pinning, cancer cells invade WAT by mixing with adipocytes as shown in Figs.~\ref{fig:cellinv_sim}A and C. Without pinning, cancer cells mainly expand radially without mixing, which yields small values for $L_{inv}'$ and $A_{inv}'$ at large $N_c/N_{c0}$ (Figs.~\ref{fig:cellinv_sim}B and D). In 2D, the main phenotype is collective cancer cell migration without adipocyte mixing, regardless of tissue pressure and cancer cell-cell adhesion (Fig.~\ref{fig:cellinv_sim}B). In contrast, this invasion phenotype only occurs for small $P$ and strong cancer cell-cell adhesion in 3D (Fig.~\ref{fig:cellinv_sim}D). Notably, for the invasion phenotype where cancer cells mix with adipocytes in 3D, we find that it takes significantly more cancer cells to reach the same level of mixing at low pinning strength $K_{pin}$ compared to high pinning strength $K_{pin}$ (Figs.~\ref{fig:cell_model_sim} C and D).

\subsection*{Tumor-derived soluble factors induce adipocyte lipid loss and dedifferentiation} 

\begin{figure*}[t!]
    \centering
    \includegraphics[width=\linewidth]{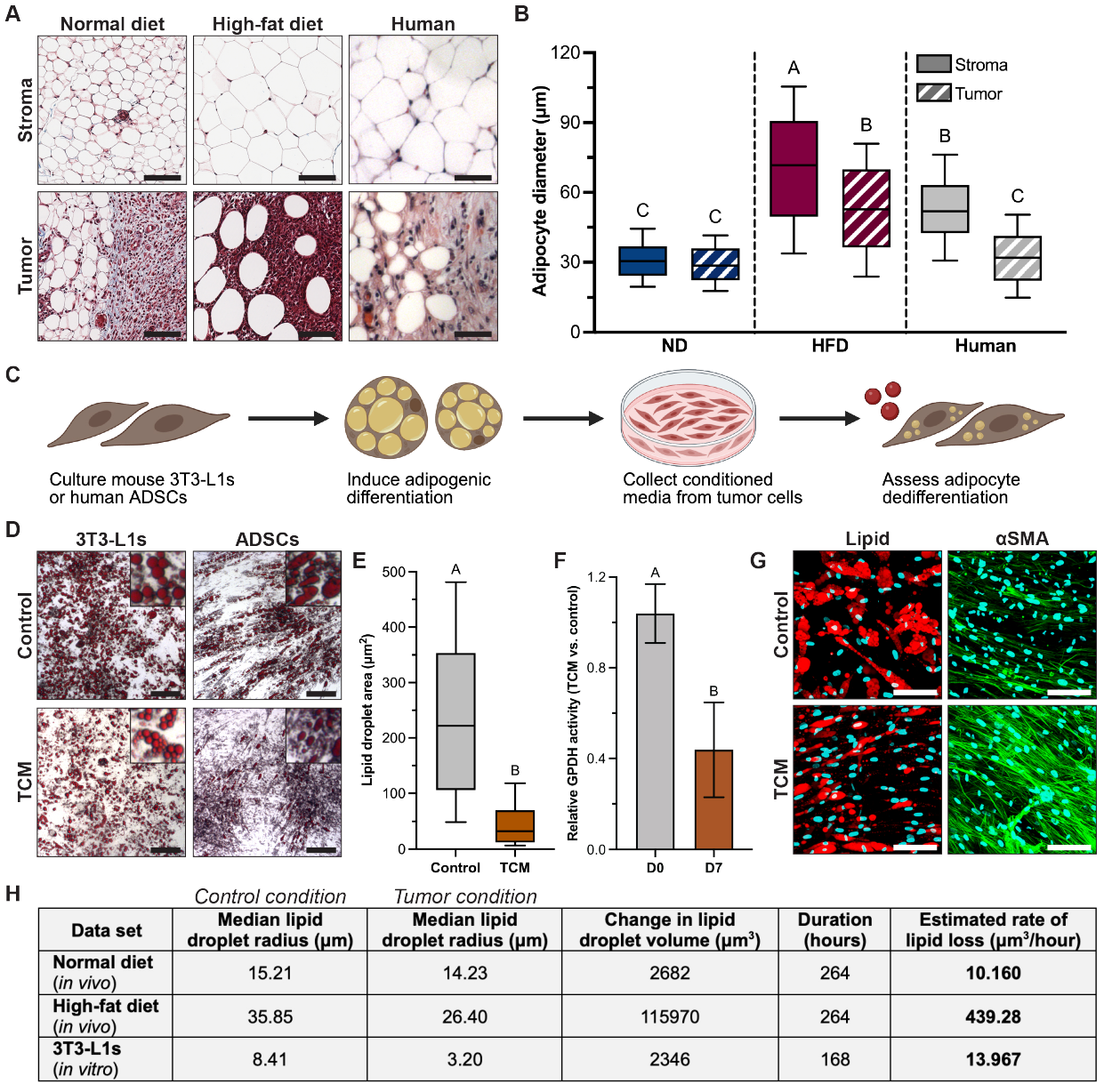}
    \caption{\textbf{Tumor-derived factors induce adipocyte lipid loss and dedifferentiation.} (A) Representative sections of adipocytes in mammary WAT (stroma) and at the tumor border (tumor) in mice pre-fed a normal diet, mice pre-fed a high-fat diet, and human samples. Scale bars = $100~\mu m$. (B) Box plot distributions of adipocyte diameters in the stroma (solid) or at the tumor border (striped). (C) Representative images of differentiated 3T3-L1 or ADSC adipocytes in 2D culture after $7$ days of treatment with control media or tumor-conditioned media (TCM). Scale bars = $200~\mu m$, insets = $100~\mu m$. (D) Box plot distributions of lipid droplet areas in differentiated 3T3-L1 adipocytes after $7$ days of treatment with control media or TCM. (E) Relative GPDH activity in differentiated 3T3-L1 adipocytes treated with TCM versus control media at Day $0$ (D0) and Day $7$ (D7). (F) Representative images of differentiated 3T3-L1 adipocytes in 3D culture stained for neutral lipid (red) and $\alpha$SMA (green) after $7$ days of treatment with control media or TCM. Scale bars = $100~\mu m$. (G) Estimated rates of lipid loss in micrometers cubed per hour from data presented in panels (B) and (D). All data are presented as mean $\pm$ standard deviation unless otherwise noted. Statistics were performed with a one-way analysis of variance (ANOVA) with Tukey's multiple comparisons test (B) or Mann-Whitney U test (E, F). Compact letter display indicates statistical significance. Groups not sharing a letter are significantly different ($p < 0.05$).}
    \label{fig:lipidloss_exp}
\end{figure*}

\begin{figure*}[t!]
    \centering
    \includegraphics[width=\linewidth]{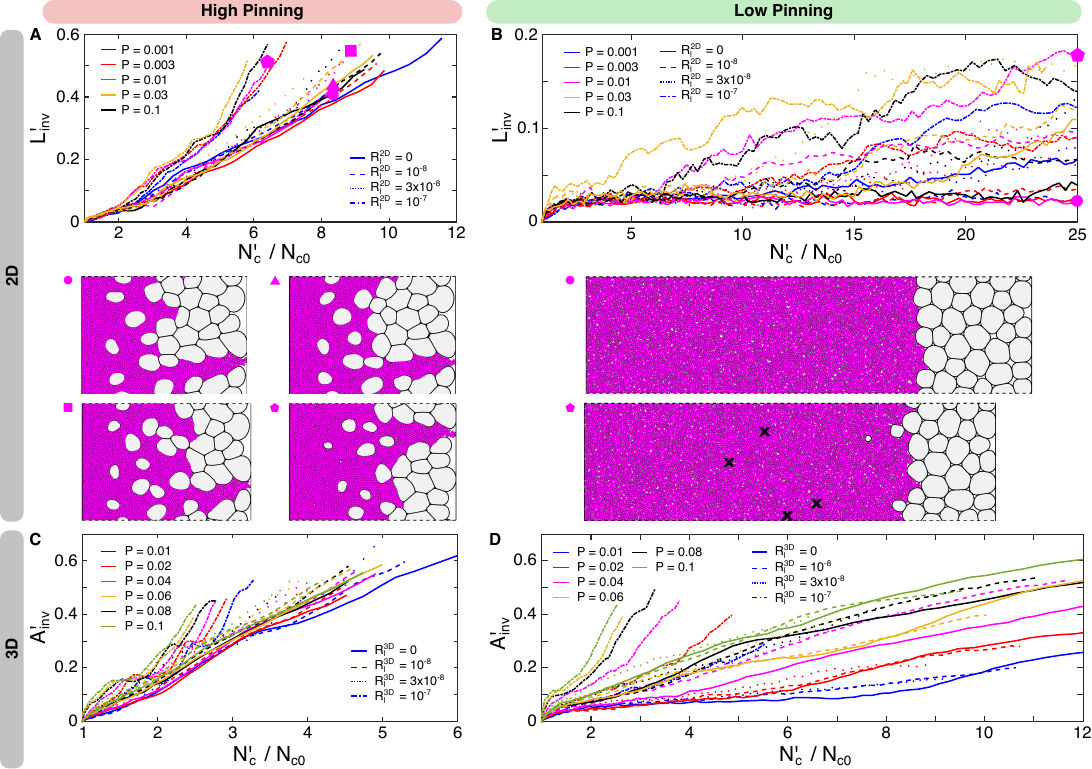}
    \caption{\textbf{DEM simulations indicate that tumor-induced lipid loss in adipocytes physically promotes invasion into WAT by increasing interstitial void space.} (A)-(B) Breast cancer invasion with lipid loss in adipocytes in 2D with cancer cell-cell adhesion parameter $a_c = 1.2$ and $\zeta = 0.2$ from Eq.~\ref{eq:energy_cc} and two values of the adipocyte pinning strength: (A) $K_{pin} = 0.2$ and (B) $K_{pin} = 0$. Top: Degree of cancer invasion $L_{inv}'$ plotted versus the number of rescaled cancer cells, $N_c'/N_{c0}$, for several values of tissue pressure $P$ and lipid loss rate $R_l^{2D}$. Bottom: Snapshots from the DEM simulations of cancer invasion into WAT with $P = 0.01$ as shown in the top panel at the end of simulations for different values of $R_l^{2D}$. Cancer cells are magenta and adipocytes are gray. Black crosses in the bottom images of panel (B) indicate extremely small adipocytes. (C)-(D) Degree of cancer invasion $A_{inv}'$ plotted versus the rescaled number of cancer cells $N_c'/N_{c0}$ in 3D with $a_c = 1.2$ and $\zeta = 0.2$ and two adipocyte pinning strengths: (C) $K_{pin} = 0.2$ and (D) $K_{pin} = 0$. DEM simulations with different tissue pressures $P$ and lipid loss rates $R_l^{3D}$ are shown by the indicated colors and symbols.}
    \label{fig:lipidloss_sim}
\end{figure*}

Given that the initial DEM simulations demonstrated that tissue pressure has little impact on tumor invasion, we next explored alternative mechanisms that could influence cancer cell invasion. Prior studies indicate that tumor-derived factors can induce adipocytes to lose their stored lipid and dedifferentiate into myofibroblast-like cells~\cite{dirat2011cancerresearch, bochet2013cancerresearch, ilina2020natcellbio}. As a result, we postulated that this tumor-induced lipid loss could alter the physical features of adipocytes, reduce the mechanical barrier imposed by tumor-adjacent WAT, and facilitate subsequent invasion. To investigate this hypothesis, we directly compared adipocytes in regions of breast tissue far from (stroma) and near (tumor) the tumor border in both the mouse and human data (Fig.~\ref{fig:lipidloss_exp}A). While we found no significant difference in adipocyte size between the stroma and tumor regions for ND mice, we observed significant decreases in adipocyte size for tumor regions in both HFD mice and human samples (Fig.~\ref{fig:lipidloss_exp}B). As the decrease in adipocyte size likely resulted from the loss of intracellular lipid, we next aimed to confirm this observation more directly using in vitro differentiated adipocytes (Fig.~\ref{fig:lipidloss_exp}C). Accordingly, we induced adipogenic differentiation in preadipocytes from mice (3T3-L1s) and humans (ADSCs) over $14$ days. In parallel, we collected and concentrated soluble factors from the human triple-negative breast cancer cell line MDA-MB-231 to generate tumor-conditioned media (TCM). TCM was then applied to fully differentiated adipocytes every other day to assess lipid loss and dedifferentiation. After $7$ days, we observed a significant decrease in the coverage of intracellular lipid droplets in adipocytes treated with TCM versus control media (Fig.~\ref{fig:lipidloss_exp}D-E). We also found that TCM treatment significantly reduced glycerol-3-phosphate dehydrogenase (GPDH) activity, which is a common marker for adipocytes and required for the production and storage of lipid in the form of triglycerides (Fig.~\ref{fig:lipidloss_exp}F). Finally, we repeated this experiment in 3D collagen hydrogels to better mimic the native environment of WAT. Similar to the results in 2D, we found that TCM treatment reduced the coverage of intracellular lipid droplets in 3D differentiated 3T3-L1 adipocytes and also promoted upregulation of the myofibroblast marker $\alpha$ smooth muscle actin ($\alpha$SMA) (Fig.~\ref{fig:lipidloss_exp}G). Collectively, these results indicate that tumor-derived soluble factors induce adipocyte lipid loss, impair canonical adipocyte function, and promote the emergence of a myofibroblast-like phenotype. We provide estimates of the lipid loss rates across these experimental measurements in Fig.~\ref{fig:lipidloss_exp}H for comparison to the DEM simulations. To obtain these estimates, the median lipid droplet radius was calculated for each adipocyte in the tumor versus stroma for \textit{in vivo} data and each differentiated adipocyte in control versus TCM conditions for \textit{in vitro} data. The difference in median lipid droplet volume was then divided by the duration of exposure to tumor-derived factors to obtain a lipid loss rate in micrometers cubed per hour.

\subsection*{Tumor-induced lipid loss in adipocytes physically promotes breast cancer invasion} 

Next, we included tumor-induced lipid loss in adipocytes during invasion in the DEM simulations to investigate how the lipid loss rate impacts the cancer invasion speed and phenotype. Notably, we find that tumor invasion is faster when adipocytes lose lipid, even when breast cancer cell proliferation and migration speed are kept constant (Fig.~\ref{fig:lipidloss_sim}). In 2D, with adipocyte pinning $K_{pin}$ and a given set of $P$, $a_c$, and $\zeta$, we find that cancer cells invade more rapidly with increasing lipid loss rate $R_l^{2D}$ (Fig.~\ref{fig:lipidloss_sim}A) due to the creation of additional void space (Fig.~\ref{fig:lipidloss_sim}A). We also find that for a given $R_l^{2D}$, cancer invasion occurs more rapidly at higher tissue pressures $P$ (Fig.~\ref{fig:lipidloss_sim}A), as the increased tissue pressure pushes proliferating cancer cells into the void spaces created by lipid loss. Without adipocyte pinning, we show that the main invasion phenotype is collective growth without mixing (Fig.~\ref{fig:lipidloss_sim}B), i.e. $L_{inv}'$ remains small even at large rescaled numbers of cancer cells $N_c'/N_{c0}$. Interestingly, we find that the typical invasion phenotype with lipid loss at high $K_{pin}$ resembles the invasion phenotype observed in HFD mice (Fig.~\ref{fig:lipidloss_exp}A), while the typical invasion phenotype with lipid loss at zero $K_{pin}$ resembles the invasion phenotype in ND mice (Fig.~\ref{fig:lipidloss_exp}A). This result suggests that adipocyte pinning to the ECM may be altered in obesity, consistent with reported increases in collagen VI, an ECM ligand for adipocyte adhesion, in obese WAT \cite{abdennour2014jcem, seo2015scitransmed}. In 3D, with adipocyte pinning $K_{pin}$, for a given set of $a_c$ and $\zeta$, we find that cancer invasion also occurs more rapidly with increasing lipid loss rate $R_l^{3D}$ at fixed $P$ and higher $P$ at fixed $R_l^{3D}$ (Fig.~\ref{fig:lipidloss_sim}C). Without adipocyte pinning in 3D, for conditions that would otherwise promote collective growth without mixing (low $P$ and high $a_c$ and $\zeta$), we find that increasing $R_l^{3D}$ changes the invasion phenotype to enable mixing of cancer cells and adipocytes (Fig.~\ref{fig:lipidloss_sim}D). The difference between 2D and 3D at $K_{pin} = 0$ is likely due to confinement, as it is easier for cancer cells to move around adipocytes when there is extra space created by lipid loss in 3D. We note that $R_l^{3D} = 10^{-8}$ and $R_l^{3D} = 10^{-7}$ correspond to lipid loss rates of approximately $20~\mu m^3/h$ and $200~\mu m^3/h$, using an average number of cancer cells in contact with an adipocyte of $N_{\alpha, c} = 10$  (Eq.~\ref{eq:lipid_loss_3d}), which are comparable to the observed values in ND and HFD mice in Fig.~\ref{fig:lipidloss_exp}H.

\section*{Conclusions and future directions}

Here, we have combined experiments and DEM simulations to investigate how mammary WAT physically regulates breast cancer invasion. In experimental studies, we found that breast cancer progresses more rapidly in HFD-fed mice than in ND-fed controls and that adipocytes in HFD-fed mice are larger and exhibit greater asphericity. Similarly, both the size and asphericity of adipocytes increase with BMI in human samples. These findings highlight how obesity changes the physical properties of adipocytes across species. Consistent with previous studies \cite{dirat2011cancerresearch}, we found that tumor-derived factors induce lipid loss and dedifferentiation of adipocytes \textit{in vitro}. In addition, adipocytes were smaller in the tumor periphery than in the stroma of HFD mice, while there was no significant difference in ND mice. Estimates of lipid loss rates indicate that adipocytes from HFD mice lose lipid more rapidly than those in ND mice, in agreement with reported increases in lipid release by large, obese adipocytes~\cite{flaherty2019science, beeghly2026cellrep}. In computational studies, DEM simulations suggest that tissue pressure $P$ regulates the asphericity of adpocytes. Yet, contrary to the expectation that higher tissue pressure $P$ would impede cancer invasion, we found that $P$ had little effect on invasion when adipocytes were strongly pinned to the ECM. Without pinning, tumors can grow uniformly without mixing with adipocytes at low $P$ and high cancer cell adhesion. Alternatively, at high pinning, cancer cells can invade as individual cells and mix with adipocytes at high $P$ and low cancer cell-cell adhesion. These results highlight adipocyte pinning as a previously overlooked parameter that strongly affects the degree and phenotype of tumor invasion. We have also shown that lipid loss physically promotes cancer invasion, even when cancer cell proliferation and migration speed are kept constant, and can also synergize with increased tissue pressure in the DEM simulations. Together, these experimental and computational studies reveal that tumor-induced lipid loss is an important process that physically remodels the local microenvironment by increasing interstitial void space and reducing the mechanical barrier to breast cancer invasion.

Moving forward, this work raises several interesting research directions. Here, we have focused on adipocyte size and asphericity, as well as adipocyte pinning to the ECM, as the physical parameters of interest. However, prior studies have shown that the stiffness of WAT also changes in obese conditions~\cite{abdennour2014jcem}. Since cancer cell migration depends on substrate stiffness~\cite{ansardamavandi03092018}, future work should quantify the relative contributions of adipocyte and ECM mechanics~\cite{abdennour2014jcem} to WAT stiffness and associated changes in tumor invasion. Second, in our current DEM simulations, cancer invasion is driven by tumor cell proliferation. However, active cell motility is also an important factor driving tumor invasion. Therefore, it will be necessary to investigate how the addition of cancer cell activity influences invasion into WAT. Moreover, breast cancer cells can metabolize the lipid lost by adipocytes to fuel their proliferation and migration~\cite{dirat2011cancerresearch, balaban2017cancermetab, beeghly2026cellrep}. While we already observe that lipid loss generates interstitial void space and reduces the mechanical barrier to invasion, directly accounting for this biochemical communication in the DEM simulations would likely enhance the pro-invasive effect of lipid loss. Finally, the DEM simulations indicate that adipocyte pinning to the surrounding ECM plays a crucial role in determining cancer invasion phenotype. These results should be validated \textit{in vitro} through our recently developed model system composed of adipose-mimetic granular hydrogel particles and interstitial collagen fibers~\cite{knode2026cellbiomater}. Collectively, the findings from these and future studies can be applied to identify the regions of WAT most susceptible to tumor invasion and target physical biomarkers to suppress the spread of breast cancer in patients.

\section*{Methods}

\subsection*{Intraductal mouse model of breast cancer}

For animal studies, female C57BL/6 mice (The Jackson Laboratories) were randomized into groups at 8–10 weeks of age and fed a high-fat diet (Research Diets \#D12492) or normal diet (LabDiet \#5053) ad libitum for 11 weeks. Diets were administered weekly and remaining chow from the previous week was discarded. Subsequently, $28,000$ ES272 or ES278 cells were injected into the ducts of the fourth mammary glands of each mouse bilaterally. ES272 cells are a syngeneic C57BL/6 breast cancer cell line with constitutively active and mutated PI3KCA (H1047R), while ES278 cells also overexpress MYC via a WAP-Cre transgene (both gifted from Dr. Ramon Parsons, Mount Sinai). Eleven days post injection, high-fat diet-fed animals had reached humane endpoints. Animals were weighed and final tumor volumes were measured with calipers. Mice were euthanized via \ce{CO2} inhalation and tumors were fixed in $4\%$ (w/v) paraformaldehyde in 1x PBS for $18$ hours at $4^{{\circ}}$C. Resected tumors were then stored in $70\%$ (v/v) ethanol in water at $4^{{\circ}}$C until processing. Samples were sent to the Cornell College of Veterinary Medicine Animal Health Diagnostic Center for paraffin embedding, sectioning, and staining with hematoxylin and eosin or Masson’s trichrome. Stained sections were imaged and digitized on an Aperio ScanScope CS2 (Leica) with a 40x objective. All animal protocols were approved by the Institutional Animal Care and Use Committees at Weill Cornell Medicine and Cornell University.

\subsection*{Collection of human breast adipose tissue}

The Institutional Review Boards of Memorial Sloan Kettering Cancer Center (IRB 10-040) and Weill Cornell Medicine (IRB 1004010984-01) approved the collection of non-tumor breast tissue under a biospecimen acquisition protocol. Tumor-containing breast tissue was collected under a separate protocol from Weill Cornell Medicine (IRB 0408007390). Informed consent was obtained from all participants. Samples were collected from patients receiving mastectomy for breast cancer risk reduction or treatment. Height and weight were recorded prior to surgery and used to calculate body mass index (BMI). Standard definitions were used to categorize BMI as normal weight (BMI $\le 25$), overweight ($25 < {\rm BMI} < 30$), or obese (BMI $\ge 30$). All data were reviewed for accuracy independently by research staff and a physician. On the day of mastectomy, paraffin blocks were prepared from resected breast tissue and all samples were de-identified before use. Two hematoxylin- and eosin-stained sections were generated from formalin-fixed, paraffin-embedded tissue to measure the diameters of adipocyte cross-sections as previously described~\cite{cho2021carcinogenesis}. The sections were photographed with a 20x objective using an Olympus BX43 or BX50 microscope equipped with an Olympus DP27 or MicroFire (Optronics) digital camera, respectively.

\subsection*{Maintenance and differentiation of preadipocytes}

Mouse 3T3-L1 preadipocytes (ATCC) were cultured in $\alpha$MEM (Sigma-Aldrich \#M4526) supplemented with $10\%$ (v/v) fetal bovine serum (FBS, R\&D Systems \#S11150) and $1\%$ (v/v) penicillin-streptomycin (ThermoFisher \#15070063). Human adipose-derived stromal cells (ADSCs, Lonza) were cultured in ADSC growth media (Lonza \#PT-4505). Both cell lines were cultured until $90\%$ confluent prior to induction of adipogenic differentiation. Adipogenic differentiation of 3T3-L1 preadipocytes was initiated by treatment with $1$ mM dexamethasone (Sigma-Aldrich \#D4902), $500$ mM IBMX (Krackeler Scientific \#45-I7018), $20$ mM indomethacin (Sigma-Aldrich \#405268), and $200$ $\mu$M insulin (Krackeler Scientific \#45-91077C). Adipogenic differentiation of ADSCs was initiated by culture in PGM-2 growth media (Lonza \#PT-8002) according to manufacturer's protocols. Differentiation media was replaced every $48$ hours until experimental endpoints were reached.

\subsection*{Collection of tumor-conditioned media}

Tumor-conditioned media (TCM) was collected as follows. The human triple-negative breast cancer cell line MDA-MB-231 (ATCC) was cultured in $\alpha$MEM supplemented with $10\%$ (v/v) FBS and $1\%$ (v/v) penicillin-streptomycin for $24$ hours and the media containing tumor-derived factors was collected. For the control medium, cell-free $\alpha$MEM supplemented with $10\%$ (v/v) FBS and $1\%$ (v/v) penicillin-streptomycin was collected in the same manner. The media was then normalized to total cell number, concentrated through a centrifugal filter unit with a molecular weight cut-off of $3$ kDa (Millipore \#UFC9003), and reconstituted with fresh $\alpha$MEM containing $10\%$ (v/v) FBS and $1\%$ (v/v) penicillin-streptomycin to make a 2-fold concentrate of tumor-derived factors for subsequent use.

\subsection*{Analysis of adipogenic differentiation}

Lipid accumulation in differentiated 3T3-L1s and ADSCs was visually assessed and quantified by Oil Red O (Millipore-Sigma \#O0625) staining as previously described. Cells were fixed in $4\%$ (w/v) paraformaldehyde in 1x PBS and stained with $0.33\%$ (w/v) Oil Red O in isopropanol for $2$ hours. Excess Oil Red O was washed from samples with 1x PBS, and the samples were imaged with a brightfield microscope (Zeiss Observer Z.1). In addition, adipogenic differentiation was assessed via glycerol-3-phosphate dehydrogenase (GPDH) activity. Cell culture supernatants were collected in a buffer containing 50 mM Tris (VWR \# 1185-53-1), 1mM ethylenediaminetetraacetic acid (EDTA, VWR \# 899301), and 1mM $\beta$-mercaptoethanol (VWR \# 60-24-2). Subsequently, the supernatants were mixed with dihydroxyacetone phosphate (Millipore Sigma \#57-04-5) and oxidized nicotinamide adenine dinucleotide (NADH, Avantor \# 53-84-9) on ice. The subsequent decrease in NADH absorbance at $340$ nm was measured on a spectrophotometer over a $7$-minute period. The GPDH activity was normalized to total protein content as measured by a BCA assay according to manufacturer’s protocols (ThermoFisher \#23225).

\subsection*{3D differentiation of preadipocytes}

For analyzing dedifferentiation of adipocytes in 3D, we microfabricated adipocyte-embedded collagen gels. The fabrication procedure was adopted from our previously published protocol. Briefly, collagen was collected from rat-tail tendons, solubilized in $0.1\%$ (v/v) acetic acid, lyophilized, and then reconstituted back in acetic acid at a final concentration of $15$ mg/mL. Then, $4$-mm-diameter and $250$-$\mu$m-thick polydimethylsiloxane (PDMS) microwells were fabricated using SYLGARD 184 silicone elastomer (Ellsworth Adhesives \#4019862) and their surfaces were treated with $1\%$ (v/v) poly(ethylenimine) (Sigma-Aldrich \#181978) and then $0.1\%$ (v/v) glutaraldehyde (ThermoFisher \#ICN19859580) to adhere collagen gels within the PDMS molds. The reconstituted collagen in acetic acid was neutralized to pH $7.2$ and then mixed with one million adipocytes to produce a final concentration of $0.6\%$ collagen. Subsequently, adipocyte-embedded collagen scaffolds were polymerized in the PDMS microwells at $37^{{\circ}}$C. The fully polymerized collagen scaffolds were cultured in either TCM or control media for $7$ days.

\subsection*{Sample fixation, permeabilization, and fluorescence staining}

Samples were fixed in cold $4\%$ (w/v) paraformaldehyde (Electron Microscopy Sciences \#19208) in 1x PBS for $15$ minutes at room temperature. Samples were washed twice with 1x PBS and permeabilized with $0.1\%$ Triton X-100 (v/v) (ThermoFisher \#A16046-AE) in 1x PBS supplemented with $1\%$ (w/v) bovine serum albumin (BSA) (ThermoFisher \#BP1600-100). Samples were washed twice with 1x PBS and then stained with Oil Red O (Millipore-Sigma \#O0625) followed by anti-alpha smooth muscle actin (anti-$\alpha$SMA, Abcam \#ab32575) in 1x PBS with $1\%$ BSA overnight at $4^{{\circ}}$C. Samples were then washed twice with 1x PBS and counterstained with goat anti-rabbit AlexaFluor 488 (Invitrogen \#A-11008) and 4',6-diamidino-2-phenylindole, dihydrochloride (DAPI, ThermoFisher \#D1306). Subsequently, the stained samples were mounted and stored with ProLong Gold (Invitrogen \# P36930).

\subsection*{Confocal microscopy and image analysis}

Images were acquired on an inverted or upright Zeiss LSM 880 confocal microscope using either a 10x/0.45 W C-Apochromat objective, 20x/0.5 EC Plan-Neofluar objective, 20x/1.0 W Plan-Apochromat objective, 40x/1.2 W C-Apochromat objective, or 63x/1.4 O Plan-Apochromat objective. All images analysis was performed in QuPath~\cite{bankhead2017qupath} and ImageJ (National Institutes of Health). For semi-automated quantification of adipocytes and lipid droplets, the AdipoQ ImageJ plug-in~\cite{sieckmann2022adipq} was used. 

\subsection*{Statistical analysis}

All experiments were performed with at least three biological replicates unless otherwise noted. Data with two conditions were evaluated with a nonparametric Mann-Whitney U test unless otherwise noted. Data with three or more conditions were evaluated with either a standard or nested one-way ANOVA with Tukey multiple comparisons test unless otherwise noted. P-values less than $0.05$ were considered statistically significant. Unless otherwise noted, all data are plotted as means with standard deviations. Compact letter display was used to show which group means are statistically different: groups sharing at least one letter are not significantly different, while groups with no letters in common are significantly different from each other. All statistical analysis was performed using GraphPad Prism (v.10.2.3) or R (v.4.3.1).

\subsection*{Shape and interaction potential energy functions for adipocytes and cancer cells}

\begin{figure*}[t!]
    \centering
    \includegraphics[width=\linewidth]{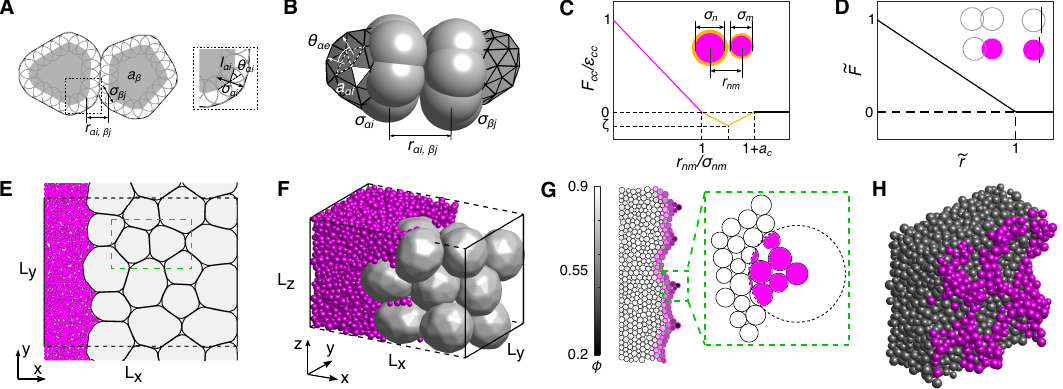}
    \caption{\textbf{Discrete element method model of breast cancer invasion into WAT in two and three dimensions.} (A) Adipocytes are modeled as deformable polygons in 2D, whose area, perimeter, and local curvature are controlled by the potential energy function $U_{a}^{2D}$ in Eq.~\ref{eq:energy_adip_2d}. Definitions of the area $a_{\beta}$, the edge length $l_{\alpha i}$, the bending angle $\theta_{\alpha i}$, vertex diameters $\sigma_{\alpha i}$ and $\sigma_{\beta j}$, and the inter-vertex separation $r_{\alpha i, \beta j}$ are included. (B) Adipocytes are modeled as deformable polyhedra in 3D, whose volume, surface area, and local curvature are controlled by the potential energy function $U_{a}^{3D}$ in Eq.~\ref{eq:energy_adip_3d}. Definitions of the surface triangle area $a_{\alpha f}$, the bending angle $\theta_{\alpha e}$, vertex diameters $\sigma_{\alpha i}$ and $\sigma_{\beta j}$, and the inter-vertex separation $r_{\alpha i, \beta j}$ are included. Several spherical vertices are omitted for clarity. (C) Inter-cellular force law between two cancer cells in both 2D and 3D from Eq.~\ref{eq:energy_cc}. (D) A schematic of the purely repulsive force between adipocytes ($\widetilde{F} = F_{aa}$, $\widetilde{r} = 1 - r_{\alpha i, \beta j} / \sigma_{\alpha i, \beta j}$ from Eq.~\ref{eq:energy_aa}, overlapping gray disks), between adipocytes and cancer cells ($\widetilde{F} = F_{ac}$, $\widetilde{r} = 1 - r_{\alpha i, n} / \sigma_{\alpha i, n}$ from Eq.~\ref{eq:energy_ac}, overlap between gray and magenta disks), between adipocytes and the confining wall ($\widetilde{F} = F_{aw}$, $\widetilde{r} = 1 - 2x_{\alpha i} / \sigma_{\alpha i}$ or $\widetilde{r} = 1 - 2(L_x -x_{\alpha i}) / \sigma_{\alpha i}$ from Eq.~\ref{eq:energy_w}, overlap between gray disk and vertical line), and between cancer cells and the confining wall ($\widetilde{F} = F_{cw}$, $\widetilde{r} = 1 - 2x_n / \sigma_n$ or $\widetilde{r} = 1 - 2(L_x -x_n) / \sigma_n$ from Eq.~\ref{eq:energy_w}, overlap between magenta disk and vertical line). (E)-(F) Illustration of cancer cells (left, magenta spheres) and adipocytes (right, gray deformable cells) with associated Cartesian coordinate axes in (E) 2D with box size $L_x \times L_y$ and (F) 3D with box dimensions $L_x \times L_y \times L_z$. Solid (dashed) lines in (E) and (F) indicate the confining walls (periodic boundaries). (G) Local packing fraction $\phi$ near each cancer cell, indicated by the gray scale, and cancer cells at the interface between cancer 
    cells and adipocytes, highlighted in magenta in 2D. (H) Cancer cells at the interface between cancer cells and adipocytes highlighted in magenta in 3D.}
    \label{fig:cell_model_sim}
\end{figure*} 

For DEM simulations of cancer invasion into WAT, we model adipocytes as deformable polygons (polyhedra) and cancer cells as adhesive soft disks (spheres) in 2D (3D). The total potential energy of the system is given by
\begin{equation}
    \label{eq:energy_total}
    U = U_{a}^{2D(3D)} + U_{aa} + U_{cc} + U_{ac} + U_{a, pin} + U_w,
\end{equation}
where $U_{a}^{2D(3D)}$ is the total shape potential energy of adipocytes in 2D (3D), $U_{aa}$ is the total interaction energy between adipocytes, $U_{cc}$ is the total interaction energy between cancer cells, $U_{ac}$ is the total interaction energy between adipocytes and cancer cells, $U_{a,pin}$ is the potential energy that pins the center of mass of the adipocytes to their equilibrium positions, and $U_w$ is the total interaction energy between both adipocytes and cancer cells and the confining walls. In 2D, each adipocyte is modeled as a deformable polygon with $N_v$ circular vertices, and $U_{a}^{2D}$ is given by
\begin{equation}
    \label{eq:energy_adip_2d}
    \begin{split}
        U_{a}^{2D} = & \sum_{\alpha}^{N_a} \frac{1}{2} \epsilon_a^{2D} \left(1 - \frac{a_\alpha}{a_{\alpha,0}}\right)^2 + \sum_{\alpha}^{N_a} \sum_{i}^{N_v} \frac{1}{2} \epsilon_l^{2D} \left(1 - \frac{l_{\alpha i}}{l_{\alpha i,0}}\right)^2 + \\
        & \sum_{\alpha}^{N_a} \sum_{i}^{N_v} \frac{1}{2} \epsilon_b^{2D} \left(\theta_{\alpha i} - \theta_{\alpha i,0}\right)^2,
    \end{split}
\end{equation}
where $N_\alpha$ is the total number of adipocytes, $a_\alpha$ is the area of adipocyte $\alpha$ with a preferred area $a_{\alpha,0}$, $l_{\alpha i}$ is the length between vertices $i$ and $(i+1)$ with a preferred length $l_{\alpha i,0}$, and $\theta_{\alpha i}$ is the angle formed by vertices $(i-1)$, $i$, and $(i+1)$ with a preferred angle $\theta_{\alpha i,0}$. $\epsilon_a$, $\epsilon_l$, and $\epsilon_b$ are the energy scales that control deviations of adipocyte area, perimeter, and local curvature from their equilibrium values (Fig.~\ref{fig:cell_model_sim}A). In 3D, each adipocyte is modeled as a deformable polyhedron tessellated by $N_f$ triangles with $N_e$ edges and $N_v$ vertices, and $U_{a}^{3D}$ is given by
\begin{equation}
    \label{eq:energy_adip_3d}
    \begin{split}
        U_{a}^{3D} = & \sum_{\alpha}^{N_a} \frac{1}{2} \epsilon_v^{3D} \left(1 - \frac{v_\alpha}{v_{\alpha,0}}\right)^2 + \sum_{\alpha}^{N_a} \sum_{f}^{N_f} \frac{1}{2} \epsilon_a^{3D} \left(1 - \frac{a_{\alpha f}}{a_{\alpha f,0}}\right)^2 + \\
        & \sum_{\alpha}^{N_a} \sum_{e}^{N_e} \frac{1}{2} \epsilon_b^{3D} \left(\theta_{\alpha e} - \theta_{\alpha e,0}\right)^2,
    \end{split}
\end{equation}
where $v_\alpha$ is the volume of adipocyte $\alpha$ with a preferred volume $v_{\alpha,0}$, $a_{\alpha f}$ is the area of triangle $f$ on adipocyte $\alpha$ with a preferred area $a_{\alpha f,0}$, and $\theta_{\alpha e}$ is the angle formed by adjacent triangles that share edge $e$ on adipocyte $\alpha$ with a preferred angle $\theta_{\alpha e,0}$ (Fig.~\ref{fig:cell_model_sim}B). To prevent adipocytes from inter-penetrating, we include a potential energy penalty when two adipocytes overlap:
\begin{equation}
    \label{eq:energy_aa}
    U_{aa} = \sum_{\alpha}^{N_a} \sum_{\beta > \alpha}^{N_a} \sum_{i}^{N_v} \sum_{j}^{N_v} \frac{1}{2} \epsilon_{aa} \left(1 - \frac{r_{\alpha i,\beta j}}{\sigma_{\alpha i,\beta j}}\right)^2 \Theta\left(1 - \frac{r_{\alpha i,\beta j}}{\sigma_{\alpha i,\beta j}}\right),
\end{equation}
where $r_{\alpha i,\beta j}$ is the distance between vertex $i$ on adipocyte $\alpha$ and vertex $j$ on adipocyte $\beta$, $\sigma_{\alpha i,\beta j}$ is the sum of the radii of vertex $i$ on adipocyte $\alpha$ and vertex $j$ on adipocyte $\beta$, and $\Theta\left(\cdot\right)$ is the Heaviside step function (Fig.~\ref{fig:cell_model_sim}B). In addition, the $N_c$ cancer cells interact pairwise via the following potential energy:
\begin{equation}
    \label{eq:energy_cc}
    U_{cc} = 
    \begin{cases}
        \sum_{n}^{N_c} \sum_{m>n}^{N_c} \frac{1}{2} \epsilon_{cc} \left(\left(1 - \frac{r_{nm}}{\sigma_{nm}}\right)^2 - \frac{1}{2} \zeta a_c^2\right), & \\
        \hfill r_{nm} \le \sigma_{nm}
        \\
        \sum_{n}^{N_c} \sum_{m>n}^{N_c} \frac{1}{2} \zeta \epsilon_{cc} \left(\left(1 - \frac{r_{nm}}{\sigma_{nm}}\right)^2 - \frac{1}{2} a_c^2\right), & \\
        \hfill \sigma_{nm} < r_{nm} \le (1 + 0.5 a_c) \sigma_{nm}
        \\
        \sum_{n}^{N_c} \sum_{m>n}^{N_c} -\frac{1}{2} \zeta \epsilon_{cc} \left(1 + a_c - \frac{r_{nm}}{\sigma_{nm}}\right)^2, & \\
        \hfill (1 + 0.5 a_c) \sigma_{nm} < r_{nm} \le (1 + a_c) \sigma_{nm}
        \\
        0, \hfill r_{nm} > (1 + a_c) \sigma_{nm}
    \end{cases}
\end{equation}
where $r_{nm}$ is the distance between cancer cells $n$ and $m$, $\sigma_{nm}$ is the sum of the radii of cancer cells $n$ and $m$, $a_{c}$ sets the attraction range, and $\zeta$ sets the attraction strength (Fig.~\ref{fig:cell_model_sim}C). The corresponding intercellular forces are illustrated in Fig.~\ref{fig:cell_model_sim}C. Adipocytes and cancer cells interact via purely repulsive forces obtained from the gradient of the following potential energy:
\begin{equation}
    \label{eq:energy_ac}
    U_{ac} = \sum_{\alpha}^{N_a} \sum_{i}^{N_v} \sum_{n}^{N_c}{\frac{1}{2} \epsilon_{ac} \left(1 - \frac{r_{\alpha i,n}}{\sigma_{\alpha i,n}}\right)^2 \Theta\left(1 - \frac{r_{\alpha i,n}}{\sigma_{\alpha i,n}}\right)},
\end{equation}
where $r_{\alpha i,n}$ is the distance between vertex $i$ on adipocyte $\alpha$ and cancer cell $n$, and $\sigma_{\alpha i,n}$ is the sum of the radii of vertex $i$ on adipocyte $\alpha$ and cancer cell $n$. To mimic tethering of adipocytes to the ECM, the centers of mass of the adipocytes are pinned via the following potential energy: 
\begin{equation}
    \label{eq:energy_pin}
    \begin{split}
        U_{a, pin} = & \sum_{\alpha}^{N_a} \frac{1}{2} k_{pin} \left( \frac{1}{N_v} \sum_{i}^{N_v} \vec{r}_{\alpha i}  - \vec{r}_{\alpha, 0} \right)^2,
    \end{split}
\end{equation}
where $\vec{r}_{\alpha, 0}$ is the equilibrium position of the center of mass for adipocyte $\alpha$. Packings of adipocytes and cancer cells are confined within a rectangular box (prism) with area $L_x L_y$ (volume $L_x L_y L_z$) and periodic boundary conditions in the $y$-direction ($y$- and $z$- directions) and confined by two walls in the $x$-direction in 2D (3D) (Figs.~\ref{fig:cell_model_sim} D and E). One wall is fixed at $x = 0$, and the other is positioned at $x = L_x$. Both adipocytes and cancer cells interact with the walls via purely repulsive forces that are generated via the gradient of the following potential energy:
\begin{equation}
    \label{eq:energy_w}
    \begin{split}
        U_w = & \sum_{\alpha}^{N_a} \sum_{i}^{N_v} \frac{1}{2}\epsilon_{aw} \left(1 - \frac{x_{\alpha i}}{0.5 \sigma_{\alpha i}}\right)^2 \Theta\left(1 - \frac{x_{\alpha i }}{0.5 \sigma_{\alpha i}}\right) + \\
        & \sum_{\alpha}^{N_a} \sum_{i}^{N_v} \frac{1}{2}\epsilon_{aw} \left(1 - \frac{L_x - x_{\alpha i}}{0.5 \sigma_{\alpha i}}\right)^2 \Theta\left(1 - \frac{L_x - x_{\alpha i }}{0.5 \sigma_{\alpha i}}\right) + \\
        & \sum_{n}^{N_c} \frac{1}{2} \epsilon_{cw} \left(1 - \frac{x_{n}}{0.5 \sigma_n}\right)^2 \Theta\left(1 - \frac{x_{n}}{0.5 \sigma_n}\right) + \\
        & \sum_{n}^{N_c} \frac{1}{2} \epsilon_{cw} \left(1 - \frac{L_x - x_{n}}{0.5 \sigma_n}\right)^2 \Theta\left(1 - \frac{L_x - x_{n}}{0.5 \sigma_n}\right),
    \end{split}
\end{equation}
where $x_{\alpha i}$ and $x_{n}$ are the x positions of vertex $i$ on adipocyte $\alpha$ and cancer cell $n$ to the closest wall, respectively, and $\sigma_{\alpha i}$ and $\sigma_n$ are diameters of vertex $i$ on adipocyte $\alpha$ and the $n$th cancer cell, respectively. The pressure $P$ in the system is given by the sum of all forces exerted on the wall at $L_x$, divided by $L_y$ ($L_yL_z$) in 2D (3D). We use $P$ as a control variable to report the tissue pressure from increased lipid content in adipocytes.

\subsection*{Demixed static packings of adipocytes and cancer cells}

For DEM simulations of cancer cell invasion into WAT, we first generate a static packing of adipocytes and cancer cells with a sharp interface at a target pressure $P_t$. To prevent adipocytes from mixing with cancer cells in the static packing, we first generate the adipocyte packing and cancer cell packing separately and then join them together.

To generate adipocyte packings in 2D, we start from a dilute system with packing fraction $\phi = 0.01$ in a square box (with lengths $L_{x,a}=L_{y,a}$ along the $x$- and $y$-directions) with two straight walls to confine the packing in the $x$-direction and periodic boundary conditions in the $y$-direction. We then compress the packing by reducing $L_{x,a}$ and $L_{y,a}$ by the same amount that gives the packing fraction increment $\Delta \phi = 0.001$, followed by energy minimization. If $P$ is smaller than $P_t$, we compress the packing again followed by energy minimization. Otherwise, we return to the packing before the compression step and compress the system by half of the original $\Delta \phi$. We repeat this process until we reach $P_t$ to within $P_t \pm 0.01P_t$. We apply the same procedure to generate a packing of breast cancer cells in a box with length in the $y$-direction $L_{y,c} = L_{y,a}$, and we only change the box length along the $x$-direction $L_{x,c}$ to achieve $P_t$. We then join the adipocyte and breast cancer cell packings and repeat the compression and decompression procedure by changing only the box length along the $x$-direction until $P$ of the entire packing reaches $P_t$ to within $P_t \pm 0.01P_t$.

In 3D, we follow the same procedure to generate the initial adiopcyte-cancer packing as in 2D, with periodic boundary conditions applied in the $y$- and $z$-directions and two confining walls located at $x = 0$ and $L_x$. We first generate an adipocyte packing in a cubic box ($L_{x,a} = L_{y,a} = L_{z,a}$). Then, we generate a cancer packing in a box with fixed sizes $L_{y,c} = L_{z,c} = L_{y,a} = L_{z,a}$ and only vary $L_{x,c}$ to achieve the target pressure. Finally, we generate the adipocyte-cancer packing with $L_{y} = L_{z} = L_{y,a} = L_{z,a}$ and only vary $L_{x}$ to achieve the target pressure.

\subsection*{Equations of motion for breast cancer invasion into adipose tissue}

After generating a demixed, static packing of adipocytes and cancer cells, we simulate breast cancer invasion into the surrounding WAT driven by cancer cell proliferation. In particular, proliferation occurs only for cancer cells in the periphery of the tumor, which is quantified by the local cancer packing fraction $\phi_n$ for cancer cell $n$. To calculate $\phi_n$, we draw a test circle (sphere) centered at cancer cell $n$ with diameter $\sigma_n + 2\left\langle D_c \right\rangle$, where $\left\langle D_c \right\rangle$ is the mean cancer cell diameter, and determine $\phi_n$ to be the ratio of the total area (volume) of the enclosed cancer cells divided by the area (volume) of the test circle (sphere) in 2D (3D). An illustration of the calculation of $\phi_n$ in 2D is shown in Fig.~\ref{fig:cell_model_sim}G. We define cancer cells with $\phi_n < \phi_c^{2D}$ ($\phi_n < \phi_c^{3D}$) to be in the periphery of the tumor, with $\phi_c^{2D} = 0.8$ in 2D (Fig.~\ref{fig:cell_model_sim}G) and $\phi_c^{3D} = 0.4$ in 3D (Fig.~\ref{fig:cell_model_sim}H). We note that the values of $\phi_c^{2D}$ and $\phi_c^{3D}$ will determine the thickness of the cancer cell-adipocyte interface, but will not change the results for cancer invasion qualitatively. Cancer cells defined to be at the interface will proliferate with a division time drawn from a Gamma distribution~\cite{chao2019molsysbio, gross2023natcomm}, with a mean $T_d$ and standard deviation $0.4T_d$. After a cancer cell divides, the daughter cells will only divide if they are in the periphery of the tumor. 

The equations of motion for the cancer cells, adipocytes, and the mobile wall are given by:
\begin{equation}
    m_{\alpha i} \frac{d^2{\vec{r}}_{\alpha i}}{dt^2} = -\frac{\partial U}{\partial{\vec{r}}_{\alpha i}} - \gamma \frac{d \vec{r}_{\alpha i}}{dt},
\end{equation}
\begin{equation}
    m_n \frac{d^2{\vec{r}}_n}{dt^2} = -\frac{\partial U}{\partial{\vec{r}}_n} - \gamma \frac{d{\vec{r}}_n}{dt},
\end{equation}
\begin{equation}
    m_w \frac{d^2 L_x}{dt^2} = -F_t \hat{x} - \frac{\partial U}{\partial{L_x}} - \gamma \frac{d L_x}{dt},
\end{equation}
where $m_{\alpha i}$, $m_n$, and $m_w$ are the masses of vertex $i$ on adipocyte $\alpha$, cancer cell $n$, and the mobile wall, respectively, $\vec{r}_{\alpha i}$ and $\vec{r}_n$ are the positions of vertex $i$ on adipocyte $\alpha$ and cancer cell $n$, $\gamma$ is the damping coefficient, $\hat{x}$ is the unit vector along positive $x$-direction, and $F_t$ is the magnitude of the external force on the mobile wall to maintain the target pressure with $F_t = P_t L_y$ ($F_t = P_t L_y L_z$) in 2D (3D). We terminate the simulations when any cancer cell reaches the wall that originally confined the adipocytes. We note that cancer proliferation during invasion will generate locally higher pressures than $P_t$ at the tumor-adipocyte interface in the limit of large pinning strength. Thus, large pinning can create a pressure gradient extending from the tumor-adipocyte interface to the boundary that confines the adipose tissue.

We also include cancer-induced lipid loss among adipocytes in the DEM simulations. Lipid loss is modeled by the following equations of motion for the equilibrium adipocyte area in 2D and volume in 3D, respectively,
\begin{equation}
    \label{eq:lipid_loss_2d}
    \frac{d a_{\alpha,0}}{dt} = -R_l^{2D} N_{\alpha, c},
\end{equation}
\begin{equation}
    \label{eq:lipid_loss_3d}
    \frac{d v_{\alpha,0}}{dt} = -R_l^{3D} N_{\alpha, c},
\end{equation}
where $R_l^{2D}$ and $R_l^{3D}$ are the lipid loss rate in 2D and 3D, and $N_{\alpha, c}$ is the number of cancer cells in contact with adipocyte $\alpha$. For non-zero $R_l^{2D}$ and $R_l^{3D}$, at any time step, we account for the required cancer cells $N_{ca}$ to fill the area and volume lost by adipocytes by reporting the adjusted cancer cell number $N_c' = N_c - N_{ca}$.

\subsection*{Characterization of the degree of cancer invasion}

\begin{figure}[t!]
    \centering
    \includegraphics[width=\linewidth]{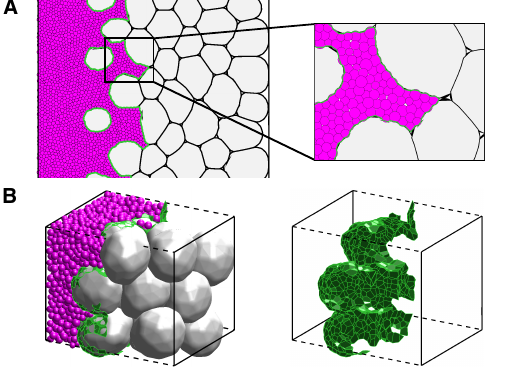}
    \caption{\textbf{Characterization of the adipocyte-cancer cell interface.} (A) Left: The adipocyte-cancer cell interface (green) is defined by the shared Voronoi edges between vertices on adipocytes (gray) and cancer cells (magenta) in 2D. Right: Close-up of the left panel to highlight the adipocyte-cancer cell interface. (B) Left: The adipocyte-cancer interface (green) is defined by the shared Voronoi polygons between vertices on adipocytes (gray) and cancer cells (magenta) in 3D. Right: The same packing as in the left panel showing only the adipocyte-cancer interface.}
    \label{fig:inv_interface}
\end{figure}

We quantify the degree of cancer cell invasion by calculating the interface length in 2D and area in 3D between adipocytes and cancer cells. We first calculate the Voronoi tessellation using the positions and diameters of the cancer cells and vertices of the adipocytes~\cite{voro++}. After Voronoi tessellation, each breast cancer cell and each adipocyte vertex are associated with their Voronoi polygon (polyhedron) in 2D (3D). All edges (polygons) are shared between two Voronoi polygons (polyhedra). We discard edges (polygons) shared between vertices that belong to the same adipocyte. In 2D, we define the degree of invasion at a given time as
\begin{equation}
    L_{inv} = \frac{L_{ac}}{L_{ac} + L_{aa} + L_{aw}},
\end{equation}
where $L_{ac}$, $L_{aa}$, and $L_{aw}$ are the total lengths of edges shared between adipocytes and cancer cells, between different adipocytes, and between adipocytes and walls, respectively. An example of an adipocyte-cancer cell interface in 2D is shown in Fig.~\ref{fig:inv_interface}A. Similarly, in 3D, we define the degree of invasion at a given time as
\begin{equation}
    A_{inv} = \frac{A_{ac}}{A_{ac} + A_{aa} + A_{aw}},
\end{equation}
where $A_{ac}$, $A_{aa}$, and $A_{aw}$ are the total areas of polygons shared between adipocytes and cancer cells, between different adipocytes, and between adipocytes and walls, respectively. An example of an adipocyte-cancer cell interface in 3D is shown in Fig.~\ref{fig:inv_interface}B.

We further rescale $L_{inv}$ ($A_{inv}$) by the minimum $L_{min}$ ($A_{min}$) and maximum $L_{max}$ ($A_{max}$) of $L_{inv}$ ($A_{inv}$) over the full course of the invasion simulations so that the rescaled degree of invasion exists between $0$ and $1$:
\begin{equation}
    L_{inv}' = \frac{L_{inv} - L_{min}}{L_{max} - L_{min}},
\end{equation}
in 2D and 
\begin{equation}
    A_{inv}' = \frac{A_{inv} - A_{min}}{A_{max} - A_{min}}
\end{equation}
in 3D. 

\subsection*{DEM simulation parameters}

In 2D, we study systems with $N_\alpha = 32$ adipocytes.  Half possess $N_v = 20$ vertices and the other half possess $N_v = 28$ vertices. The system includes $N_c = N_{c0} = 300$ cancer cells at the beginning of simulations. For each adipocyte, we set the preferred asphericity in 2D $\mathcal{A}^{2D} = (N_v l_{\alpha,0})^2 / (4 \pi a_{\alpha,0}) = 1$ and $\theta_{\alpha,0} = 2 \pi / N_v$, such that a stress-free adipocyte possesses the shape of a regular polygon with $N_v$ sides. In 3D, we study systems with $N_\alpha = 32$ adipocytes possessing $N_v = 100$ vertices and $N_f = 196$ surface triangles and $N_c = N_{c0} = 1500$ cancer cells at the beginning of simulations. For each adipocyte, we set the vertices of a stress-free adipocyte to be at the equilibrium positions of electrons from the Thomson problem with $N_v$ electrons~\cite{wales2006prb}, which allows us to calculate $a_{\alpha f,0}$, $v_{\alpha, 0}$, and $\theta_{\alpha e,0}$. This choice ensures that asphericity in 3D $\mathcal{A}^{3D} = (\sum_f^{N_f} a_{\alpha f,0})^{3/2} / (6 \sqrt{\pi} v_{\alpha,0}) = 1$ and that the adipocytes possess the shape of a polyhedron closest to a sphere when stress-free. We set $a_{\alpha,0} = 5000 \mu m^2$  ($10000 \mu m^2$) for the small (large) adipocytes with $N_v = 20$ $(28)$ in 2D and $v_{\alpha, 0} = 523600 \mu m^3$ so that the average adipocyte diameter is $\left\langle D_a \right\rangle = 100 \mu m$. We draw the diameters of cancer cells from a uniform distribution with mean $\left\langle D_c \right\rangle = 0.1 \left\langle D_a \right\rangle = 10 \mu m$ and standard deviation $\delta_c / \left\langle D_c \right\rangle = 0.1$.

We set $\epsilon_a^{2D} = 0.05$~J and $\epsilon_v^{3D} = 5$~J so that they are comparable to the bulk modulus of a lipid droplet ($\sim 10$~MPa, $\epsilon_a^{2D} / a_{\alpha,0}$ in 2D and $\epsilon_v^{3D} / v_{\alpha,0}$ in 3D)~\cite{tat2000jaocs}. Furthermore, in 2D, we set $\epsilon_l^{2D} / \epsilon_a^{2D} = 0.5$, $\epsilon_{aa} / \epsilon_a^{2D} =  0.1$, $\epsilon_{cc} / \epsilon_a^{2D} = 0.05$, $\epsilon_{ac} / \epsilon_a^{2D} = 0.1$, $\epsilon_w / \epsilon_a^{2D} = 0.1$, and $\epsilon_b / \epsilon_a^{2D} = 10^{-4}$. In 3D, we set $\epsilon_a^{3D} / \epsilon_v^{3D} = 10$, $\epsilon_{aa} / \epsilon_v^{3D} =  0.1$, $\epsilon_{cc} / \epsilon_v^{3D} = 0.05$, $\epsilon_{ac} / \epsilon_v^{3D} = 0.1$, $\epsilon_w / \epsilon_v^{3D} = 0.1$, and $\epsilon_b / \epsilon_v^{3D} = 10^{-4}$. We set $T_d / \sqrt{\left\langle D_c \right\rangle^2 m_c / \epsilon_{cc}} = 24$ hours~\cite{schiffer1979cancer}. We report $P$ in units of $\epsilon_a / a_{\alpha, 0}$ ($\epsilon_v / v_{\alpha, 0}$) in 2D (3D), $K_{pin}$ in units of $\epsilon_a / a_{\alpha, 0}^2$ ($\epsilon_v / v_{\alpha, 0}^2$) in 2D (3D), $R_l^{2D}$ in units of $a_{\alpha,0} / \sqrt{\left\langle D_c \right\rangle^2 m_c / \epsilon_{cc}}$, and $R_l^{3D}$ in units of $v_{\alpha,0} / \sqrt{\left\langle D_c \right\rangle^2 m_c / \epsilon_{cc}}$.

\section*{Acknowledgments}

We acknowledge support from the National Institutes of Health under Grant No. R01CA276392 (C.S.O., C.F.), the National Science Foundation under Grant No. DGE1650441 (G.F.B.), the National Cancer Institute under Grant No. F31CA278410 (G.F.B.), the National Cancer Institute under Grant No. R01CA259195 (C.F.), Grant No. R01CA257254 (L.T.V.), and the Center on the Physics of Cancer Metabolism under Grant No. 1U54CA210184 (C.F.). This work was also supported by the High Performance Computing Facilities operated by Yale’s Center for Research Computing. We also acknowledge support from the Cornell Biotechnology Resource Center Genomics Facility and the Cornell Biotechnology Resource Center Imaging Facility funded by NYSTEM C029155, NIH S10OD018516, and NIH S10RR025502.

\section*{Author Declarations}

\subsection*{Conflict of Interest}

The authors have no conflicts to disclose.

\subsection*{Ethics Approval}

Ethics approval is not required.

\subsection*{Author Contributions}

\textbf{Garrett F. Beeghly:} Conceptualization (equal); formal analysis (equal); investigation (equal); methodology (equal); project administration (equal); software (equal); validation (equal); visualization (equal); writing – original draft (equal). \textbf{Dong Wang:} Conceptualization (equal); formal analysis (equal); investigation (equal); methodology (equal); project administration (equal); software (equal); validation (equal); visualization (equal); writing – review and editing (equal). \textbf{Bo Ri Seo:} Conceptualization (supporting); investigation (supporting); methodology (supporting). \textbf{Yitong Zheng:} Conceptualization (supporting); investigation (supporting); methodology (supporting); software (supporting). \textbf{Joseph E. Druso:} Conceptualization (supporting); investigation (supporting); methodology (supporting). \textbf{Benjamin D. Hopkins:} Conceptualization (supporting); investigation (supporting); methodology (supporting). \textbf{Linda T. Vahdat:} Conceptualization (supporting); resources (supporting); supervision (supporting). \textbf{Neil M. Iyengar:} Conceptualization (supporting); resources (supporting); supervision (supporting). \textbf{Mark D. Shattuck:} Writing – original draft (supporting); formal analysis (supporting); methodology (supporting); supervision (supporting). \textbf{Corey S. O'Hern:} Conceptualization (equal); funding acquisition (equal); resources (equal); supervision (equal); writing – original draft (equal). \textbf{Claudia Fischbach:} Conceptualization (equal); funding acquisition (equal); resources (equal); supervision (equal); writing – original draft (equal).

\section*{Data Availability}

Data from this study are available from the corresponding author upon reasonable request. The code to simulate breast cancer invasion into white adipose tissue can be found at \href{https://github.com/wangd1213/BreastCancerInvasion}{https://github.com/wangd1213/BreastCancerInvasion}.

\section*{References}

\bibliography{main}

@article{harbeck2019natreview,
    author = {Nadia Harbeck and Fr{\'e}d{\'e}rique Penault-Llorca and Javier Cortes and Michael Gnant and Nehmat Houssami and Philip Poortmans and Kathryn Ruddy and Janice Tsang and Fatima Cardoso},
    title = {Breast cancer},
    journal = {Nat. Rev. Dis. Primers},
    volume = {5},
    pages = {66},
    year = {2019},
    doi = {10.1038/s41572-019-0111-2}
}

@article{calle2003nejm,
    author = {Eugenia E. Calle  and Carmen Rodriguez  and Kimberly Walker-Thurmond  and Michael J. Thun },
    title = {Overweight, Obesity, and Mortality from Cancer in a Prospectively Studied Cohort of U.S. Adults},
    journal = {N. Engl. J. Med.},
    volume = {348},
    pages = {1625-1638},
    year = {2003},
    doi = {10.1056/NEJMoa021423}
}

@article{renehan2008lancet,
    title = {Body-mass index and incidence of cancer: a systematic review and meta-analysis of prospective observational studies},
    journal = {The Lancet},
    volume = {371},
    pages = {569-578},
    year = {2008},
    doi = {https://doi.org/10.1016/S0140-6736(08)60269-X},
    author = {Andrew G Renehan and Margaret Tyson and Matthias Egger and Richard F Heller and Marcel Zwahlen}
}

@article{protani2010,
    title = {Effect of obesity on survival of women with breast cancer: systematic review and meta-analysis},
    journal = {Breast Cancer Res. Treat.},
    volume = {123},
    pages = {627–635},
    year = {2010},
    doi = {10.1007/s10549-010-0990-0},
    author = {Melinda Protani and Michael Coory and Jennifer H. Martin}
}

@article{park2011endorev,
    author = {Park, Jiyoung and Euhus, David M. and Scherer, Philipp E.},
    title = {Paracrine and Endocrine Effects of Adipose Tissue on Cancer Development and Progression},
    journal = {Endocr. Rev.},
    volume = {32},
    pages = {550-570},
    year = {2011},
    doi = {10.1210/er.2010-0030}
}

@article{khandekar2011natrevcancer,
    title = {Molecular mechanisms of cancer development in obesity},
    journal = {Nat. Rev. Cancer},
    volume = {11},
    pages = {886–895},
    year = {2011},
    doi = {10.1038/nrc3174},
    author = {Melin J. Khandekar and Paul Cohen and Bruce M. Spiegelman}
}

@article{hopkins2016jco,
    title = {Obesity and Cancer Mechanisms: Cancer Metabolism},
    journal = {J. Clin. Oncol.},
    volume = {34},
    pages = {4277–4283},
    year = {2016},
    doi = {10.1200/JCO.2016.67.9712},
    author = {Benjamin D Hopkins and Marcus D Goncalves and Lewis C Cantley}
}

@article{ilina2020natcellbio,
    title = {Cell--cell adhesion and 3D matrix confinement determine jamming transitions in breast cancer invasion},
    author = {Ilina, Olga and Gritsenko, Pavlo G and Syga, Simon and Lippoldt, J{\"u}rgen and La Porta, Caterina AM and Chepizhko, Oleksandr and Grosser, Steffen and Vullings, Manon and Bakker, Gert-Jan and Starru{\ss}, J{\"o}rn and others},
    journal = {Nat. Cell. Biol.},
    volume = {22},
    number = {9},
    pages = {1103--1115},
    year = {2020},
    doi = {10.1038/s41556-020-0552-6}
}

@article{kleinert2018natrevendo,
    title = {Animal models of obesity and diabetes mellitus},
    author = {Maximilian Kleinert and Christoffer Clemmensen and Susanna M. Hofmann and Mary C. Moore and Simone Renner and Stephen C. Woods and Peter Huypens, Johannes Beckers and Martin Hrabe de Angelis and Annette Sch{\"u}rmann and Mostafa Bakhti and Martin Klingenspor and Mark Heiman and Alan D. Cherrington and Michael Ristow and Heiko Lickert and Eckhard Wolf and Peter J. Havel and Timo D. M{\"u}ller and Matthias H. Tsch{\"o}p},
    journal = {Nat. Rev. Endocrinol.},
    volume = {14},
    pages = {140--162},
    year = {2018},
    doi = {10.1038/nrendo.2017.161}
}

@article{cross2007natnanotechnol,
    title = {Nanomechanical analysis of cells from cancer patients},
    author = {Cross, Sarah E and Jin, Yu-Sheng and Rao, Jianyu and Gimzewski, James K},
    journal = {Nat. Nanotechnol.},
    volume = {2},
    pages = {780–783},
    year = {2007},
    doi = {10.1038/nnano.2007.388}
}

@article{rianna2020mbc,
    title = {Direct evidence that tumor cells soften when navigating confined spaces},
    author = {Rianna, Carmela and Radmacher, Manfred and Kumar, Sanjay},
    journal = {Mol. Biol. Cell},
    volume = {31},
    number = {16},
    pages = {1726--1734},
    year = {2020},
    doi = {10.1091/mbc.E19-10-0588}
}

@article{provenzano2006bmcmed,
    title = {Collagen reorganization at the tumor-stromal interface facilitates local invasion},
    author = {Provenzano, Paolo P and Eliceiri, Kevin W and Campbell, Jay M and Inman, David R and White, John G and Keely, Patricia J},
    journal = {BMC Med.},
    volume = {4},
    pages = {38},
    year = {2006},
    doi = {10.1186/1741-7015-4-38}
}

@article{levental2009cell,
    title = {Matrix Crosslinking Forces Tumor Progression by Enhancing Integrin Signaling},
    journal = {Cell},
    volume = {139},
    number = {5},
    pages = {891-906},
    year = {2009},
    doi = {https://doi.org/10.1016/j.cell.2009.10.027},
    url = {https://www.sciencedirect.com/science/article/pii/S0092867409013531},
    author = {Kandice R. Levental and Hongmei Yu and Laura Kass and Johnathon N. Lakins and Mikala Egeblad and Janine T. Erler and Sheri F.T. Fong and Katalin Csiszar and Amato Giaccia and Wolfgang Weninger and Mitsuo Yamauchi and David L. Gasser and Valerie M. Weaver}
}

@article{sakers2022cell,
    title = {Adipose-tissue plasticity in health and disease},
    journal = {Cell},
    volume = {185},
    number = {3},
    pages = {419-446},
    year = {2022},
    issn = {0092-8674},
    doi = {https://doi.org/10.1016/j.cell.2021.12.016},
    url = {https://www.sciencedirect.com/science/article/pii/S0092867421014549},
    author = {Alexander Sakers and Mirian Krystel {De Siqueira} and Patrick Seale and Claudio J. Villanueva}
}

@article{ghaben2019natrevmcb,
    title = {Adipogenesis and metabolic health},
    author = {Ghaben, A. L. and Scherer, P. E.},
    journal = {Nat. Rev. Mol. Cell Biol.},
    volume = {20},
    pages = {242–-258},
    year = {2019},
    doi = {10.1038/s41580-018-0093-z}
}

@article{seo2015scitransmed,
    author = {Bo Ri Seo and Priya Bhardwaj and Siyoung Choi and Jacqueline Gonzalez and Roberto C. Andresen Eguiluz and Karin Wang and Sunish Mohanan and Patrick G. Morris and Baoheng Du and Xi K. Zhou and Linda T. Vahdat and Akanksha Verma and Olivier Elemento  and Clifford A. Hudis and Rebecca M. Williams and Delphine Gourdon and Andrew J. Dannenberg and Claudia Fischbach},
    title = {Obesity-dependent changes in interstitial ECM mechanics promote breast tumorigenesis},
    journal = {Sci. Transl. Med.},
    volume = {7},
    number = {301},
    pages = {301ra130-301ra130},
    year = {2015},
    doi = {10.1126/scitranslmed.3010467},
    URL = {https://www.science.org/doi/abs/10.1126/scitranslmed.3010467}
}

@article{dirat2011cancerresearch,
    author = {Dirat, B{\'e}atrice and Bochet, Ludivine and Dabek, Marta and Daviaud, Dani{\`e}le and Dauvillier, St{\'e}phanie and Majed, Bilal and Wang, Yuan Yuan and Meulle, Aline and Salles, Bernard and Le Gonidec, Sophie and Garrido, Ignacio and Escourrou, Ghislaine and Valet, Philippe and Muller, Catherine},
    title = {Cancer-Associated Adipocytes Exhibit an Activated Phenotype and Contribute to Breast Cancer Invasion},
    journal = {Cancer Res.},
    volume = {71},
    number = {7},
    pages = {2455-2465},
    year = {2011},
    doi = {10.1158/0008-5472.CAN-10-3323}
}

@article{bochet2013cancerresearch,
    author = {Ludivine Bochet and Camille Lehu{\'e}d{\'e} and St{\'e}phanie Dauvillier and Yuan Yuan Wang and B{\'e}atrice Dirat and Victor Laurent and Cédric Dray and Romain Guiet and Isabelle Maridonneau-Parini and Sophie Le Gonidec and Bettina Couderc and Ghislaine Escourrou and Philippe Valet and Catherine Muller},
    title = {Adipocyte-Derived Fibroblasts Promote Tumor Progression and Contribute to the Desmoplastic Reaction in Breast Cancer},
    journal = {Cancer Res.},
    volume = {73},
    number = {718},
    pages = {5657–5668},
    year = {2013},
    doi = {10.1158/0008-5472.CAN-13-0530}
}

@article{naftaly2022ijms,
    author = {Alex Naftaly and Nadav Kislev and Roza Izgilov and Raizel Adler and Michal Silber and Ruth Shalgi and Dafna Benayahu},
    title = {Nutrition Alters the Stiffness of Adipose Tissue and Cell Signaling},
    journal = {Int. J. Mol. Sci.},
    volume = {23},
    number = {23},
    pages = {15237},
    year = {2022},
    doi = {10.3390/ijms232315237}
}

@article{anderson2000theormed,
    author = {Anderson, A. R. A. and Chaplain, M. A. J. and Newman, E. L. and Steele, R. J. C. and Thompson, A. M.},
    title = {Mathematical Modelling of Tumour Invasion and Metastasis},
    journal = {J. Theor. Med.},
    volume = {2},
    number = {2},
    pages = {129-154},
    doi = {https://doi.org/10.1080/10273660008833042},
    year = {2000}
}

@article{anderson2006cell,
    title = {Tumor Morphology and Phenotypic Evolution Driven by Selective Pressure from the Microenvironment},
    journal = {Cell},
    volume = {127},
    number = {5},
    pages = {905--915},
    year = {2006},
    doi = {https://doi.org/10.1016/j.cell.2006.09.042},
    url = {https://www.sciencedirect.com/science/article/pii/S0092867406013481},
    author = {Alexander R. A. Anderson and Alissa M. Weaver and Peter T. Cummings and Vito Quaranta}
}

@article{araujo2004bullmathbio,
    title = {A history of the study of solid tumour growth: The contribution of mathematical modelling},
    author = {Araujo, R. P. and McElwain, D. L. S.},
    journal = {Bull. Math. Biol.},
    volume = {66},
    pages = {1039--1091},
    year = {2004},
    doi = {10.1016/j.bulm.2003.11.002}
}

@article{bi2015naturephys,
    title = {A density-independent rigidity transition in biological tissues},
    author = {Dapeng Bi and J. H. Lopez and J. M. Schwarz and M. Lisa Manning},
    journal = {Nat. Phys.},
    volume = {11},
    pages = {1074–1079},
    year = {2015},
    doi = {10.1038/nphys3471},
    url = {https://www.nature.com/articles/nphys3471}
}

@article{bi2016prx,
    title = {Motility-Driven Glass and Jamming Transitions in Biological Tissues},
    author = {Bi, Dapeng and Yang, Xingbo and Marchetti, M. Cristina and Manning, M. Lisa},
    journal = {Phys. Rev. X},
    volume = {6},
    issue = {2},
    pages = {021011},
    numpages = {13},
    year = {2016},
    doi = {10.1103/PhysRevX.6.021011},
    url = {https://link.aps.org/doi/10.1103/PhysRevX.6.021011}
}

@article{sussman2018softmatter,
   author = {Daniel M. Sussman and Matthias Merkel},
   journal = {Soft Matter},
   pages = {3397-3403},
   title = {No unjamming transition in a Voronoi model of biological tissue},
   volume = {14},
   year = {2018},
}

@article{graner1992prl,
    title = {Simulation of biological cell sorting using a two-dimensional extended Potts model},
    author = {Graner, Fran\c{c}ois and Glazier, James A.},
    journal = {Phys. Rev. Lett.},
    volume = {69},
    issue = {13},
    pages = {2013--2016},
    year = {1992},
    doi = {10.1103/PhysRevLett.69.2013},
    url = {https://link.aps.org/doi/10.1103/PhysRevLett.69.2013}
}

@article{bresler2019epje,
    title = {Sharp interface model for elastic motile cells},
    author = {Bresler, Y. and Palmieri, B. and Grant, M.},
    journal = {Eur. Phys. J. E},
    volume = {42},
    pages = {52},
    year = {2019},
    doi = {10.1140/epje/i2019-11815-x}
}

@article{kasza2007curropiocellbio,
    title = {The cell as a material},
    journal = {Curr. Opin. Cell Biol.},
    volume = {19},
    number = {1},
    pages = {101--107},
    year = {2007},
    doi = {https://doi.org/10.1016/j.ceb.2006.12.002},
    url = {https://www.sciencedirect.com/science/article/pii/S0955067406001839},
    author = {Karen E Kasza and Amy C Rowat and Jiayu Liu and Thomas E Angelini and Clifford P Brangwynne and Gijsje H Koenderink and David A Weitz}
}

@article{boromand2018prl,
    title = {Jamming of Deformable Polygons},
    author = {Boromand, Arman and Signoriello, Alexandra and Ye, Fangfu and O'Hern, Corey S. and Shattuck, Mark D.},
    journal = {Phys. Rev. Lett.},
    volume = {121},
    issue = {24},
    pages = {248003},
    year = {2018},
    doi = {10.1103/PhysRevLett.121.248003},
    url = {https://link.aps.org/doi/10.1103/PhysRevLett.121.248003}
}

@article{treado2021prmater,
    title = {Bridging particle deformability and collective response in soft solids},
    author = {Treado, John D. and Wang, Dong and Boromand, Arman and Murrell, Michael P. and Shattuck, Mark D. and O'Hern, Corey S.},
    journal = {Phys. Rev. Mater.},
    volume = {5},
    issue = {5},
    pages = {055605},
    year = {2021},
    doi = {10.1103/PhysRevMaterials.5.055605},
    url = {https://link.aps.org/doi/10.1103/PhysRevMaterials.5.055605}
}

@article{wang2021softmatter,
   author = {Dong Wang and John D. Treado and Arman Boromand and Blake Norwick and Michael P. Murrell and Mark D. Shattuck and Corey S. O'Hern},
   issue = {43},
   journal = {Soft Matter},
   pages = {9901-9915},
   title = {The structural, vibrational, and mechanical properties of jammed packings of deformable particles in three dimensions},
   volume = {17},
   year = {2021},
}

@article{treado2022jrsi,
   author = {John D. Treado and Adam B. Roddy and Guillaume Théroux-Rancourt and Liyong Zhang and Chris Ambrose and Craig R. Brodersen and Mark D. Shattuck and Corey S. O'Hern},
   issue = {197},
   journal = {Journal of the Royal Society Interface},
   title = {Localized growth and remodelling drives spongy mesophyll morphogenesis},
   volume = {19},
   year = {2022},
   pages={20220602},
}

@article{zheng2024aplbio,
    author = {Zheng, Yitong and Wang, Dong and Beeghly, Garrett and Fischbach, Claudia and Shattuck, Mark D. and O'Hern, Corey S.},
    title = {Computational modeling of the physical features that influence breast cancer invasion into adipose tissue},
    journal = {APL Bioeng.},
    volume = {8},
    number = {3},
    pages = {036104},
    year = {2024},
    doi = {10.1063/5.0209019}
}

@article{voro++,
    author = {Chris H. Rycroft},
    title = {{VORO}++: A three-dimensional {V}oronoi cell library in {C}++},
    journal = {Chaos},
    volume = {19},
    pages = {041111},
    year = {2009},
    doi = {10.1063/1.3215722}
}

@article{tat2000jaocs,
    title = {The speed of sound and isentropic bulk modulus of biodiesel at {$21^{\circ}$C} from atmospheric pressure to 35{MPa}},
    author = {Mustafa E. Tat and Jon H. van Gerpen and Seref Soylu and Mustafa Canacki and Abdul Monyem and Samuel Wormley},
    journal = {J. Am. Oil Chem. Soc.},
    volume = {77},
    pages = {285-289},
    year = {2000},
    doi = {10.1007/s11746-000-0047-z}
}

@article{balaban2017cancermetab,
    title = {Adipocyte lipolysis links obesity to breast cancer growth: adipocyte-derived fatty acids drive breast cancer cell proliferation and migration},
    author = {Balaban, Seher and Shearer, Robert F. and Lee, Lisa S. and van Geldermalsen, Michelle and Schreuder, Mark and Shtein, Harrison C. and Cairns, Rose and Thomas, Kristen C. and Fazakerley, Daniel J. and Grewal, Thomas and Holst, Jeff and Saunders, Darren N. and Hoy, Andrew J.},
    journal = {Cancer Metab.},
    volume = {5},
    pages = {1},
    year = {2017},
    doi = {10.1186/s40170-016-0163-7}
}

@article{abdennour2014jcem,
    title = {Association of adipose tissue and liver fibrosis with tissue stiffness in morbid obesity: links with diabetes and BMI loss after gastric bypass},
    author = {Meriem Abdennour and Sophie Reggio and Gilles {Le Naour} and Yuejun Liu and Christine Poitou and Judith Aron-Wisnewsky and Frederic Charlotte and Jean-Luc Bouillot and Adriana Torcivia and Magali Sasso and Veronique Miette and Jean-Daniel Zucker and Pierre Bedossa and Joan Tordjman and Karine Clement},
    journal = {J. Clin. Endocrinol. Metab.},
    volume = {99},
    pages = {898–907},
    year = {2014},
    doi = {10.1210/jc.2013-3253}
}

@article{ansardamavandi03092018,
    author = {Arian Ansardamavandi and Mohammad Tafazzoli-Shadpour and Mohammad Ali Shokrgozar},
    title = {Behavioral remodeling of normal and cancerous epithelial cell lines with differing invasion potential induced by substrate elastic modulus},
    journal = {Cell Adh. Migr.},
    volume = {12},
    pages = {472--488},
    year = {2018},
    doi = {10.1080/19336918.2018.1475803},
    URL = {https://doi.org/10.1080/19336918.2018.1475803}
}

@article{gross2023natcomm,
    author = {Sean M. Gross and Farnaz Mohammadi and Crystal Sanchez-Aguila and Paulina J. Zhan and Tiera A. Liby and Mark A. Dane and Aaron S. Meyer and Laura M. Heiser},
    title = {Analysis and modeling of cancer drug responses using cell cycle phase-specific rate effects},
    journal = {Nat. Commun.},
    volume = {14},
    pages = {3450},
    year = {2023},
    doi = {10.1038/s41467-023-39122-z}
}

@article{chao2019molsysbio,
    author = {Hui Xiao Chao and Randy I Fakhreddin and Hristo K Shimerov and Katarzyna M Kedziora and Rashmi J Kumar and Joanna Perez and Juanita C Limas and Gavin D Grant and Jeanette Gowen Cook and Gaorav P Gupta and Jeremy E Purvis},
    title = {Evidence that the human cell cycle is a series of uncoupled, memoryless phases},
    journal = {Mol. Syst. Biol.},
    volume = {15},
    pages = {MSB188604},
    year = {2019},
    doi = {10.15252/msb.20188604}
}

@article{beeghly2022review,
   author = "Beeghly, Garrett F. and Amofa, Kwasi Y. and Fischbach, Claudia and Kumar, Sanjay",
   title = "Regulation of Tumor Invasion by the Physical Microenvironment: Lessons from Breast and Brain Cancer", 
   journal= "Annu. Rev. Biomed. Eng.",
   year = "2022",
   volume = "24",
   pages = "29-59",
   doi = "https://doi.org/10.1146/annurev-bioeng-110220-115419",
   url = "https://www.annualreviews.org/content/journals/10.1146/annurev-bioeng-110220-115419"
}

@article{beeghly2023review,
   author = "Beeghly, Garrett F. and Amofa, Kwasi Y. and Fischbach, Claudia and Kumar, Sanjay",
   title = "Measuring and modelling tumour heterogeneity across scales", 
   journal= "Nat. Rev. Bioeng.",
   year = "2023",
   volume = "1",
   pages = "712–730",
   doi = "0.1038/s44222-023-00087-9"
}

@article{knode2026cellbiomater,
    author = {Brianna K. Knode and Garrett F. Beeghly and Brittany E. Schutrum and Dong Wang and Yitong Zheng and Alice Battistella and Ruchi Goswami and Chi-Yong Eom and Irina Kopyeva and Aline Bozec and Jochen Guck and Nozomi Nishimura and Salvatore Girardo and Corey S. O’Hern and Claudia Fischbach},
    title = {Adipose-mimetic granular hydrogels uncover biophysical cues driving breast cancer invasion},
    journal = {Cell Biomater.},
    volume = {2},
    pages = {100411},
    year = {2026},
    doi = {https://doi.org/10.1016/j.celbio.2026.100411}
}

@article{wales2006prb,
  title = {Structure and dynamics of spherical crystals characterized for the Thomson problem},
  author = {Wales, David J. and Ulker, Sidika},
  journal = {Phys. Rev. B},
  volume = {74},
  pages = {212101},
  year = {2006},
  doi = {10.1103/PhysRevB.74.212101},
  url = {https://link.aps.org/doi/10.1103/PhysRevB.74.212101}
}

@article{schiffer1979cancer,
    author = {Schiffer, L. M. and Braunschweiger, P. G. and Stragand, J. J. and Poulakos, L.},
    title = {The cell kinetics of human mammary cancers},
    journal = {Cancer},
    volume = {43},
    number = {5},
    pages = {1707-1719},
    doi = {https://doi.org/10.1002/1097-0142(197905)43:5<1707::AID-CNCR2820430522>3.0.CO;2-A},
    url = {https://acsjournals.onlinelibrary.wiley.com/doi/abs/10.1002/1097-0142%28197905%2943%3A5%3C1707%3A%3AAID-CNCR2820430522%3E3.0.CO%3B2-A},
    year = {1979}
}

@article{sieckmann2022adipq,
    author = {Katharina Sieckmann and Nora Winnerling and Mylene Huebecker and Philipp Leyendecker  and Dalila Juliana Silva Ribeiro and Thorsten Gnad and Alexander Pfeifer and Dagmar Wachten and Jan N. Hansen and Diane Lidke},
    title = {Adipo{Q} - a simple, open-source software to quantify adipocyte morphology and function in tissues and in vitro},
    journal = {Mol. Biol. Cell.},
    volume = {33},
    pages = {br22},
    year = {2022},
    doi = {10.1091/mbc.E21-11-0592},
    URL = {https://www.molbiolcell.org/doi/abs/10.1091/mbc.E21-11-0592}
}

@article{bankhead2017qupath,
    author = {Peter Bankhead and Maurice B. Loughrey and Jos{\'e} A. Fern{\'a}ndez and Yvonne Dombrowski and Darragh G. McArt and Philip D. Dunne and Stephen Mc{Q}uaid and Ronan T. Gray and Liam J. Murray and Helen G. Coleman and Jacqueline A. James and Manuel Salto-Tellez and Peter W. Hamilton},
    title = {QuPath: {O}pen source software for digital pathology image analysis},
    journal = {Sci. Rep.},
    volume = {7},
    pages = {16878},
    year = {2017},
    doi = {10.1038/s41598-017-17204-5}
}

@article{beeghly2026cellrep,
    title = {Large adipocytes increase vesicle-mediated lipid release and promote breast cancer malignancy},
    journal = {Cell Rep.},
    volume = {45},
    pages = {117061},
    year = {2026},
    doi = {https://doi.org/10.1016/j.celrep.2026.117061},
    url = {https://www.sciencedirect.com/science/article/pii/S2211124726001397},
    author = {Garrett F. Beeghly and Irina Kopyeva and Jenny Deng and Marlee I. Pincus and Rohan R. Varshney and Dilip D. Giri and Domenick J. Falcone and Michael C. Rudolph and Marc A. Antonyak and Neil M. Iyengar and Claudia Fischbach}
}

@article{flaherty2019science,
    author = {Stephen E. Flaherty and Ambar Grijalva and Xiaoyuan Xu and Eleanore Ables and Alireza Nomani and Anthony W. Ferrante},
    title = {A lipase-independent pathway of lipid release and immune modulation by adipocytes},
    journal = {Science},
    volume = {363},
    number = {6430},
    pages = {989-993},
    year = {2019},
    doi = {10.1126/science.aaw2586},
    URL = {https://www.science.org/doi/abs/10.1126/science.aaw2586}
}

@article{cho2021carcinogenesis,
    author = {Cho, Byuri Angela and Iyengar, Neil M and Zhou, Xi Kathy and Morrow, Monica and Giri, Dilip D and Verma, Akanksha and Elemento, Olivier and Pollak, Michael and Dannenberg, Andrew J},
    title = {Blood biomarkers reflect the effects of obesity and inflammation on the human breast transcriptome},
    journal = {Carcinogenesis},
    volume = {42},
    number = {10},
    pages = {1281-1292},
    year = {2021},
    doi = {10.1093/carcin/bgab066},
    url = {https://doi.org/10.1093/carcin/bgab066}
}

\end{document}